\documentclass[%
reprint,
superscriptaddress,
amsmath,amssymb,
aps,
pre,
floatfix,
longbibliography
]{revtex4-2}

\usepackage{times}
\usepackage{amstext}
\usepackage{graphicx}
\usepackage{color}
\usepackage{soul}
\usepackage{MnSymbol}
\usepackage{subfiles}
\usepackage{bm}
\usepackage{hyperref}
\usepackage{layouts}
\usepackage{enumitem}
\usepackage{natbib}
\usepackage{comment}
\usepackage{xcolor}
\usepackage{booktabs}
\usepackage{longtable}
\usepackage{supertabular}
\usepackage{placeins}

\newcommand{\Vz}{\bar{V}_z}
\newcommand{\sigVz}{\delta{V_z}}
\newcommand{\Hagl}{H_{\mathrm{AGL}}}
\newcommand{\Emvz}{\mathbb{E}[\bar{V}_z]}
\newcommand{\Esvz}{\mathbb{E}[\delta{V_z}]}
\newcommand{\ttd}{T{-}T_d}
\newcommand{\vsink}{V_{\mathrm{sink}}}

\begin{document}

\title{{From Flight Logs to Atmospheric Science\\
\normalsize{Paragliders as Convection Sensors for Identifying Thermal Predictors}}}

\author{C\'esar Hern\'andez-Aguayo}

\affiliation{Econophysics Lab, Institut Louis Bachelier, 28 Pl.~de la Bourse, Palais Brongniart, 75002 Paris, France}
\affiliation{LadHyX UMR CNRS 7646, École Polytechnique, Institut Polytechnique de Paris, 91128 Palaiseau Cedex, France}

\author{Matthieu Cristelli}
\affiliation{Capital Fund Management, 23 Rue de l’Université, 75007 Paris, France}

\author{Michael Benzaquen}\email{michael.benzaquen@cfm.com}
\affiliation{Econophysics Lab, Institut Louis Bachelier, 28 Pl.~de la Bourse, Palais Brongniart, 75002 Paris, France}
\affiliation{Capital Fund Management, 23 Rue de l’Université, 75007 Paris, France}

\begin{abstract}
Atmospheric thermal convection drives boundary-layer dynamics and vertical exchange of heat, moisture, and momentum, yet fundamental questions about thermal structure remain open due to limited in situ observational coverage. We introduce a novel high-resolution observational dataset for atmospheric convection based on paragliding flight logs collected over metropolitan France during 2017–2024. Paragliders probe thermal updrafts by circling within them while carrying GPS variometers that record position and altitude; each climbing segment samples the vertical velocity field within a thermal column. Aggregated across 1.47 million climbing segments from 110,730 flights, this dataset provides unprecedented spatial coverage and temporal resolution. To demonstrate the value of this  observational resource, we extract three physically distinct observables from climbing segments and characterize their dependence on terrain, season, time of day, cloud state, and soil moisture. Coupling the paragliding observations with global atmospheric reanalysis data (ERA5, 0.25°/hourly) through a regression framework against 120 physically interpretable predictors, we identify leading atmospheric predictors of these observables. The empirical relationship between ceiling height and temperature–dew point depression matches the theoretical lifting condensation level scaling with striking precision supporting our methodology. Boundary-layer height emerges as the leading independent predictor of both ceiling height and thermal strength across all terrains and seasons, while vertical-velocity variability is controlled by surface heat-flux and wind variables. Taken together, our results highlight  the utility of crowd-sourced flight-log data for investigating atmospheric convection.
\end{abstract}

\date{\today}

\maketitle

\section{Introduction}
\label{sec:intro}
Atmospheric thermal convection, or the buoyancy-driven vertical motion of air heated by contact with a warm surface, is a fundamental process shaping the boundary layer, cloud formation, and the vertical exchange of heat, moisture, and momentum between the surface and the free atmosphere~\cite{Turner1969, Stull1988}. Since the pioneering laboratory experiments of Sparrow, Husar and Goldstein~\cite{Sparrow1970}, thermals have been described as rising plumes with a well-defined structure, periodicity, and vorticity. However, the scarcity of atmospheric observations, due to the limited availability of in situ probes, means that many fundamental questions about their spatio-temporal statistics remain open.  In particular, how are their onset, intensity, and depth controlled by atmospheric conditions, and how do these properties vary with the diurnal cycle, season, and terrain? 
Theoretical work has shown that boundary-layer depth and orography jointly influence the strength and location of thermal updrafts~\cite{KirshbaumWang2014}, but direct observational tests of these predictions remain sparse.

In situ information on thermal structure has historically come from radiosonde ascents and targeted airborne campaigns~\cite{Lenschow1980, Blyth2005}, and more recently from high-resolution large-eddy simulations {(LES)}~\cite{HernandezDeckers2016, Peters2019}. Each of these approaches remains limited in its coverage: radiosondes sample only a small number of fixed stations, typically twice daily; field campaigns are short-lived and geographically constrained; and LES requires carefully specified boundary conditions and substantial computational resources. Meanwhile, thermal-scale processes remain only partially represented in numerical weather prediction models, where they are typically treated through sub-grid parameterizations rather than being explicitly resolved~\cite{Pergaud2009, Sherwood2013}.

Beyond conventional atmospheric instruments, soaring birds have been used as natural tracers of thermal convection. Early observations of herring gulls suggested that soaring behavior reflects the intensity and structure of boundary-layer convection~\cite{Woodcock1975}, and later work on American white pelicans showed that bird trajectories can be related to thermal updraft profiles in a way broadly consistent with aircraft measurements and LES~\cite{Shannon2002}. Multi-species studies over complex terrain further indicate that birds exploit the same thermal boundary layer in different ways~\cite{ShamounBaranes2003}. Although scientifically valuable, these studies remain inherently limited in scope. They typically involve relatively small numbers of tracked animals, restricted spatial domains, and short observation periods, and they cannot deliver the kind of long-term, geographically extensive, and well controlled statistics needed to characterize thermal convection at scale. In that respect, paragliders offer a distinct opportunity: they sample the same thermals, but with flight logs recorded in a standardized way and with far denser, more systematic coverage over years, across a broad range of terrains and atmospheric conditions.

In this paper, we introduce a new high-resolution observational dataset for atmospheric thermal convection based on paragliding flight logs collected over metropolitan France. Paragliders climb in thermal updrafts by circling within them, and pilots carry onboard GPS variometers that record position and altitude throughout the flight. Therefore, each climbing segment provides a direct probe of a thermal column, sampling both the vertical structure of the updraft and the maximum height reached. Aggregated across hundreds of thousands of flights, these logs yield an observational record of thermal convection with a spatial and temporal richness that no conventional atmospheric probe network can match.

To illustrate the potential of this dataset for investigating atmospheric convection, we extract three observables from climbing segments that characterize thermal structure and use them to explore how thermals vary with terrain, season, time of day, cloud state, and soil moisture. We further couple the paragliding data with ECMWF ERA5 global atmospheric reanalysis to suggest a methodology for identifying the main atmospheric drivers of these observables, by linking thermal properties extracted from flight logs to physically interpretable atmospheric predictors.

The paper is organized as follows. Section~\ref{sec:data} describes the paragliding flight logs dataset, the ERA5 reanalysis, and the terrain classification scheme used throughout. Section~\ref{sec:methods} details the extraction of three thermal observables from climbing segments, their aggregation to cell-hour resolution, and defines the conditional splits (terrain, season, time of day, cloud cover, {and} soil moisture) used to characterize thermal variability. Section~\ref{sec:results} presents the climatological statistics across these splits, introduces the ERA5 variable-selection pipeline that reduces 206 raw variables to 120 physically interpretable predictors, validates the methodology against classical lifting-condensation-level theory, and applies a regression to identify the leading atmospheric predictors of each observable. Finally, Section~\ref{sec:conc} summarizes the main findings, discusses their physical interpretation, acknowledges limitations of the paragliding sampling, and outlines future extensions of this methodology.

\section{Data}
\label{sec:data}
\subsection{Paragliding flight logs}
The FFVL CFD~\footnote{FFVL stands for F\'ed\'eration Fran\c{c}aise de Vol Libre, and CFD for Compétition F\'ed\'erale de Distance.} is an annual  French  paragliding distance competition in which pilots submit their flight logs after each cross-country flight and are ranked by distance flown. To count, a flight generally has to be at least 15 km and start in France, and the scoring system awards points based on distance, with bonuses for triangle or out-and-return flights. The annual ranking is usually based on the pilot’s three best flights of the year, and the system relies on declared flights that must respect aviation rules. We counted 110,730 flights recorded between 2017 and 2024. The reason we do not extend the sample further back in time is to avoid bias in climbing metrics from earlier generations of wings with different performance characteristics. Each flight is provided in the IGC (International Gliding Commission) format, containing: GPS position (latitude, longitude), GNSS altitude $z$ (m above mean sea level), barometric altitude when available, and UTC timestamp sampled at approximately 1Hz.

\begin{figure}
 \centering
\includegraphics[width=0.9\columnwidth]{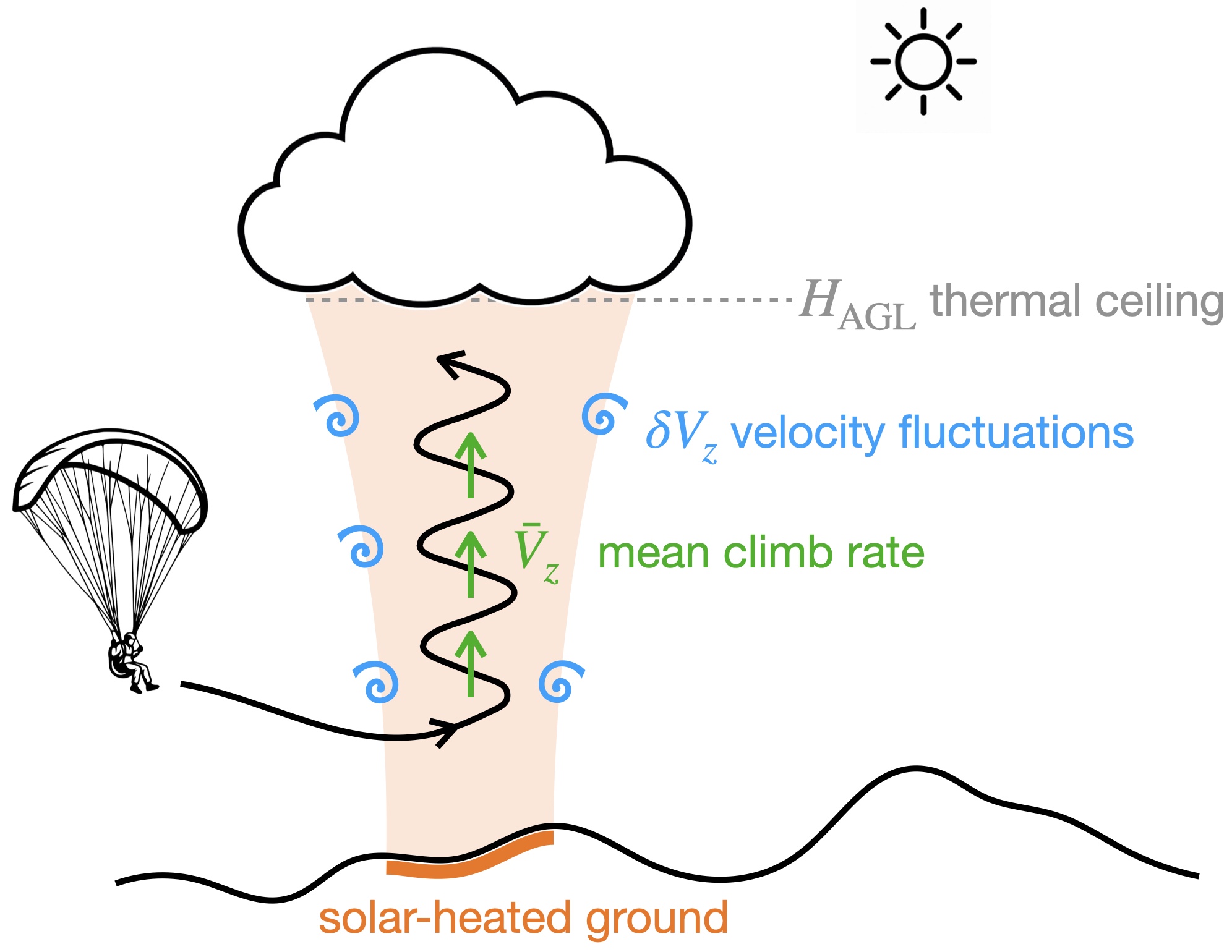}
\caption{Schematic of a paragliding climbing segment during which the pilot circles within a thermal column. The updraft rises from the heated surface up to a maximum altitude, which may correspond to either the dry or wet thermal ceiling, depending on atmospheric conditions. From such climbing segments, we extract three physical observables (see Section~\ref{sec:climbvars}):  thermal strength $\Vz$, which probes the intensity of buoyancy generation; vertical velocity variability $\sigVz$, reflecting the amplitude of  small-scale updraft fluctuations; and thermal ceiling $\Hagl$, measuring the vertical extent of convection.}
\label{fig:thermal}
\end{figure}

Flight logs are analyzed using a phase-detection algorithm~\cite{vilpellet2026} that assigns each GPS fix to one of three flight phases: \emph{search} (the pilot is scanning for lift), \emph{climb} (the pilot is circling within an updraft), and \emph{transition} (the pilot is gliding between thermals). In our analysis, we focus on the climbing segments to which we apply a set of physical quality filters: duration in the range 20-1800s (to reject spikes and anomalously long segments), altitude gain $\geq 10$m (to ensure a genuine climb), mean climb rate $\geq 0.1$ms$^{-1}$ (to exclude ambiguous or drifting phases), altitude range 50--5500m~AMSL~\footnote{Above Mean Sea Level} (to remove near-ground and stratospheric outliers), and vertical velocity within $\pm 10$ms$^{-1}$ (to remove GPS glitches). After filtering, the dataset contains 1.47 million climbing segments distributed across the French territory between 2017 and 2024. Figure~\ref{fig:thermal} illustrates a climbing phase  and the geometry of the associated thermal column. The observables extracted from these climbing segments are defined below, in Section~\ref{sec:climbvars}.

\subsection{ERA5 reanalysis}
The European Centre for Medium-Range Weather Forecasts (ECMWF) ERA5 reanalysis~\cite{ERA5:2020} is the fifth-generation global atmospheric reanalysis, combining a numerical weather model (IFS Cy41r2) with a comprehensive observation network through 4D-Var data assimilation. ERA5 provides hourly estimates of atmospheric, land, and oceanic variables on a $0.25^\circ \times 0.25^\circ$ grid, corresponding to a horizontal resolution of approximately 25km.

For this study, we extracted 206 ERA5 single-level variables over metropolitan France, defined by the bounding box 5°W–10°E, 41°–52°N, which comprises 2,745 grid cells, for the period 1990–2024. The extraction spans all relevant physical families of variables, including near-surface temperature and dew point; radiative fluxes (shortwave and longwave, both at the surface and at the top of the atmosphere, under clear-sky and all-sky conditions); turbulent fluxes (sensible heat, latent heat, and momentum); soil moisture at four depth layers; surface characteristics such as albedo and vegetation; boundary-layer variables such as layer depth and cloud base height; and vertically integrated atmospheric quantities. To remove redundancy and focus on physically independent predictors, we apply a four-stage filtering procedure (detailed in Section~\ref{sec:era5vars}) that reduces the initial 206 variables to a final set of 120 predictors used in the regression analysis.

\begin{figure}
 \centering
\includegraphics[width=\columnwidth]{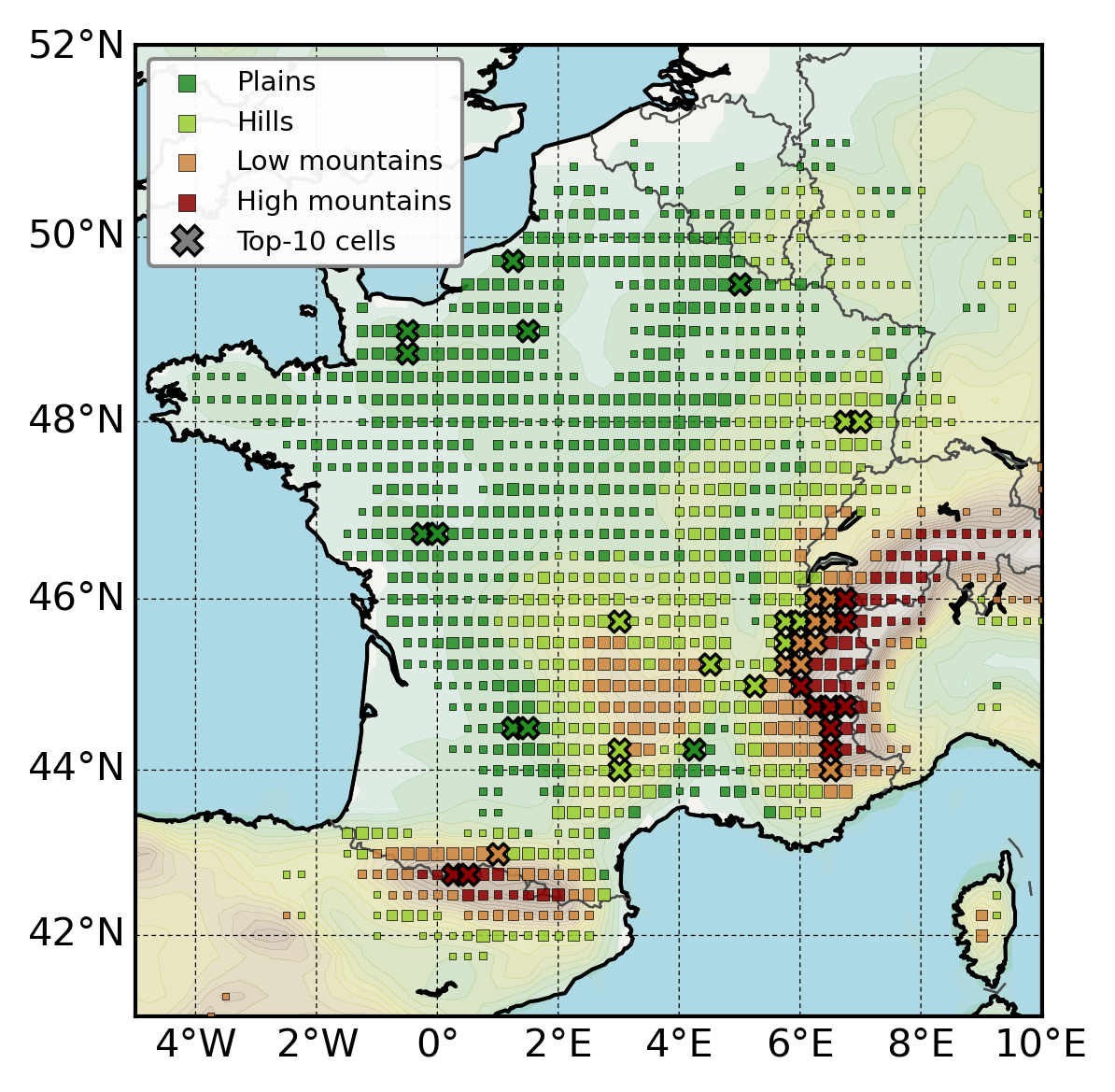}
\caption{Spatial distribution of climbing activity over France. Active ERA5 grid cells are colored by terrain class following the four-class elevation scheme of Eq.~\ref{eq:terrain}. Cell area is proportional to the logarithm of the segment count. Colored cross markers correspond to the ten highest-activity cells per terrain (forty cells total).  The background shading shows the ground elevation derived from the ERA5 surface geopotential.}
\label{fig:map}
\end{figure}

\subsection{Terrain classification}
Each ERA5 grid point is classified into one of four terrain categories based on the ERA5 surface geopotential height i.e. the ground elevation above mean sea level, $z^{\mathrm{gnd}} = \Phi/g$, where $\Phi$ is the ERA5 surface geopotential and $g=9.80665$ms$^{-2}$ is the gravitational acceleration~\footnote{The thresholds (300m, 800m, 1500m) were chosen to match the global mountain-area classification of ref.~\cite{Kapos2000}.}:
\begin{equation}
  \text{Terrain} = \begin{cases}
    \text{\,Plains}          & z^{\mathrm{gnd}} < 300\,\text{m} \\
    \text{\,Hills}           & 300 \leq z^{\mathrm{gnd}} < 800\,\text{m} \\
    \text{\,Low mountains}   & 800 \leq z^{\mathrm{gnd}} < 1500\,\text{m} \quad \\
    \text{\,High mountains}  & z^{\mathrm{gnd}} \geq 1500\,\text{m}\,.
  \end{cases}
  \label{eq:terrain}
\end{equation}

Each climbing segment is assigned to the nearest ERA5 grid cell ($0.25^\circ \times 0.25^\circ$) centered on the ERA5 grid point and matched to the corresponding hourly ERA5 time step. Coastal cells with land-sea mask $<0.9$ are excluded since flights by the coast rely primarily on dynamic lift (coastal soaring) rather than thermal convection. On the same grounds, only solar hours 9--17h are retained~\footnote{Solar time is defined by the local sun position and computed from the UTC timestamp of each IGC record using Spencer's formula~\cite{Spencer1971}, which combines the longitude correction $4\lambda$ (in minutes) with the equation of time. For a site near $\lambda = 6^\circ$E in the French Alps, solar noon lies within $\pm 16$~min of $11{:}36$~UTC over the year, i.e.\ approximately $12{:}36$ CET in winter or $13{:}36$ CEST in summer.}, as thermal activity is generally negligible outside this interval. Figure~\ref{fig:map} shows the spatial distribution of 1065 active cells over metropolitan France, with cross markers highlighting the ten top-activity cells in each terrain category. The activity is concentrated along the Alps and pre-Alps (southeastern France), with substantial contributions from the Massif Central, the Pyr\'en\'ees, and scattered flying sites in the Plains of northern and western France. The forty top-activity cells  capture 64.1\% of all climbing segments, span elevations from $\sim$100m to $>2000$m, and cover a broad range of soil, vegetation, and climatological conditions.

\section{Methodology}
\label{sec:methods}
As argued above, atmospheric thermals have been relatively poorly studied due to the lack of dense probing infrastructure. In this section, we present a methodology that uses paragliding climbing segments as probes of thermal columns and extracts physically distinct observables characterizing the statistics of individual updrafts.

\subsection{Physical observables}
\label{sec:climbvars}
From the set of climbing segments we extract three meaningful observables that characterize physical aspects of the underlying thermal column (see Fig.~\ref{fig:thermal}).\\ 

 \paragraph*{Thermal strength.} We choose the average $\Vz$ of the GPS vertical velocity  during the climbing phase as a good proxy for the intensity of buoyancy generation or thermal strength:
    \begin{equation}
        \Vz^{(i)}= \frac{1}{N_i}\sum_{j=1}^{N_i} V_{z,j}^{(i)}\,,
        \label{eq:vzbar}
    \end{equation}
where $V_{z,j}^{(i)}$ is the instantaneous vertical velocity at fix $j$ within the segment $i$, and $N_i$ is the number of fixes. Note that the GPS  vertical velocity corresponds to the motion of the aircraft, not to that of the surrounding air. Indeed, one has $V_{z} = w_{\mathrm{thermal}} - \vsink$, where $w_{\mathrm{thermal}}$ is the atmospheric updraft velocity and $\vsink \approx 1.0$--$1.2\,\mathrm{ms^{-1}}$ is the paraglider’s still-air sink rate. \\ 
    
\paragraph*{Vertical-velocity variability.} The amplitude of the vertical velocity fluctuations experienced by the pilot within the thermal $\sigVz$ is computed as the sample standard deviation of the instantaneous vertical velocity \footnote{The symbol $\delta$ is used here rather than the more customary $\sigma$ to emphasize that this quantity is a segment-level fluctuation amplitude, distinct from an ensemble variance.}, and is used here as a proxy for small-scale turbulence within the updraft~\footnote{We also considered the fourth central moment (kurtosis) of the vertical velocity within each climb as a potential additional observable of turbulence structure. However, preliminary analysis showed that this quantity has very weak Spearman correlations with all ERA5 predictors ($|\rho_S| < 0.05$ across all terrains) and does not provide statistically meaningful information at the cell-hour aggregation scale. We therefore restricted the analysis to the three observables $\Hagl$, $\Vz$, and $\sigVz$.}:
    \begin{equation}
        \sigVz^{(i)} = \sqrt{\frac{1}{N_i-1}\sum_{j=1}^{N_i} \left(V_{z,j}^{(i)} - \Vz^{(i)}\right)^2}.
        \label{eq:sigmavz}
    \end{equation}

\paragraph*{Thermal ceiling.} The maximum altitude $\Hagl$ reached above ground level (AGL), computed over all climbing segments within a given $1^\circ$ grid cell and one-hour bin, is taken as a proxy for the thermal ceiling. 
{This rests on the expectation that at least one pilot in that cell reaches the ceiling at least once, and that the thermal convection characteristics can be considered constant during that hour.}
In cell $c$  on day $d$ at hour $h$: 
    \begin{equation}
        \Hagl^{(c,d,h)} = \max_{\text{segments $i$}\in(c,d,h)} \big(z^{\mathrm{top}}_i - z^{\mathrm{gnd}}(c)\big)\,,
        \label{eq:hagl}
    \end{equation}
where $z^{\mathrm{top}}_i$ is the maximum altitude above mean sea level (AMSL) reached by the paraglider during the $i$-th climb  and $z^{\mathrm{gnd}}(c)$ is the ground elevation AMSL at the ERA5 cell $c$, derived from the surface geopotential.
Let us stress that while $\Vz$  and $\sigVz$ are defined for each climbing segment $i$ in our dataset, the ceiling $\Hagl$ is defined per cell and hour bin $(c,d,h)$. Also note that, depending on moisture conditions, the thermal ceiling may be either \emph{wet}, leading to cloud formation, or \emph{dry}, in which case no cloud forms.

\begin{figure}
 \centering
\includegraphics[width=\columnwidth]{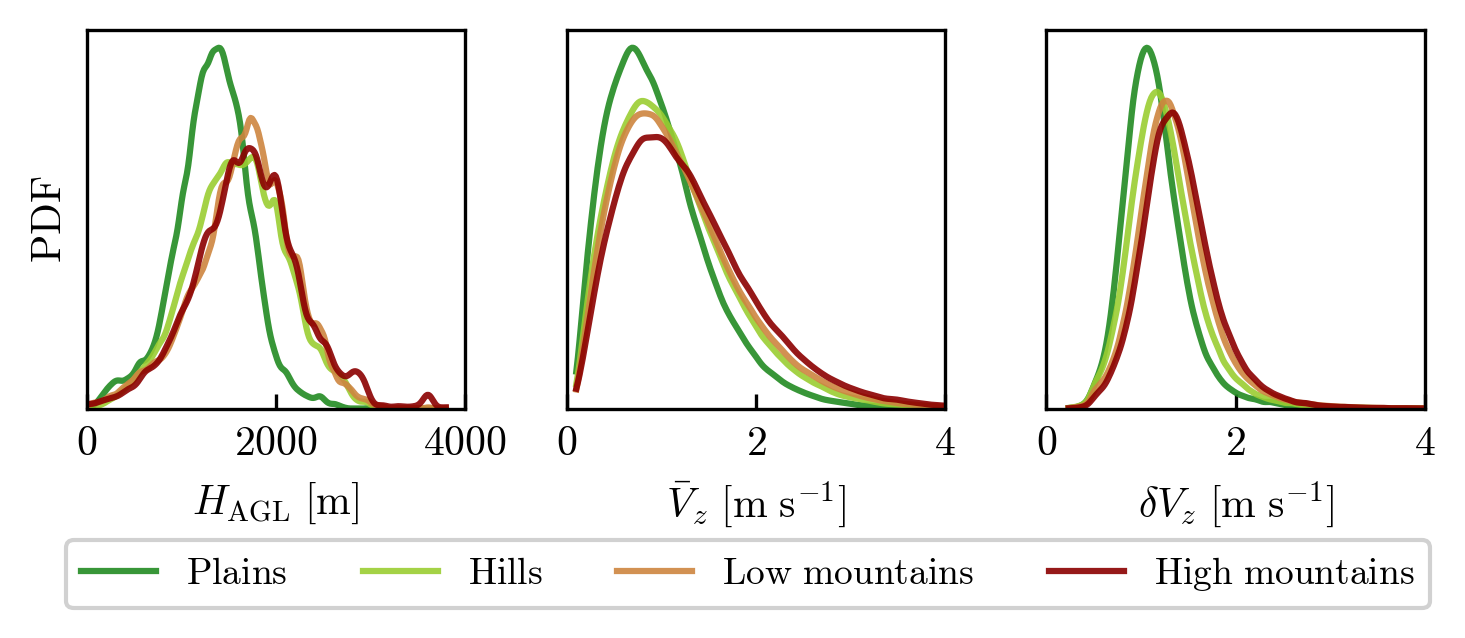}
\caption{Probability density functions of the three climbing observables per terrain class: thermal ceiling $\Hagl$ (left), thermal strength $\Vz$ (middle), and vertical velocity variability $\sigVz$ (right).}
\label{fig:pdfs}
\end{figure}

Figure~\ref{fig:pdfs} shows the empirical PDFs of the three observables across terrain classes. The thermal ceiling $\Hagl$ exhibits strong terrain dependence, with plains peaking at $\sim$1100m~AGL, while hills and mountains peak at $\sim$1800m; the heavy right tail in mountains reflects occasional deep thermals that can reach up to 3000--4000m~AGL. Thermal strength $\Vz$ shows a clear terrain gradient, with $\sim$1.0ms$^{-1}$ in Plains and $\sim$1.3ms$^{-1}$ in High mountains, while $\sigVz\sim1.0$--$1.5$ms$^{-1}$.

\begin{figure}
 \centering
\includegraphics[width=\columnwidth]{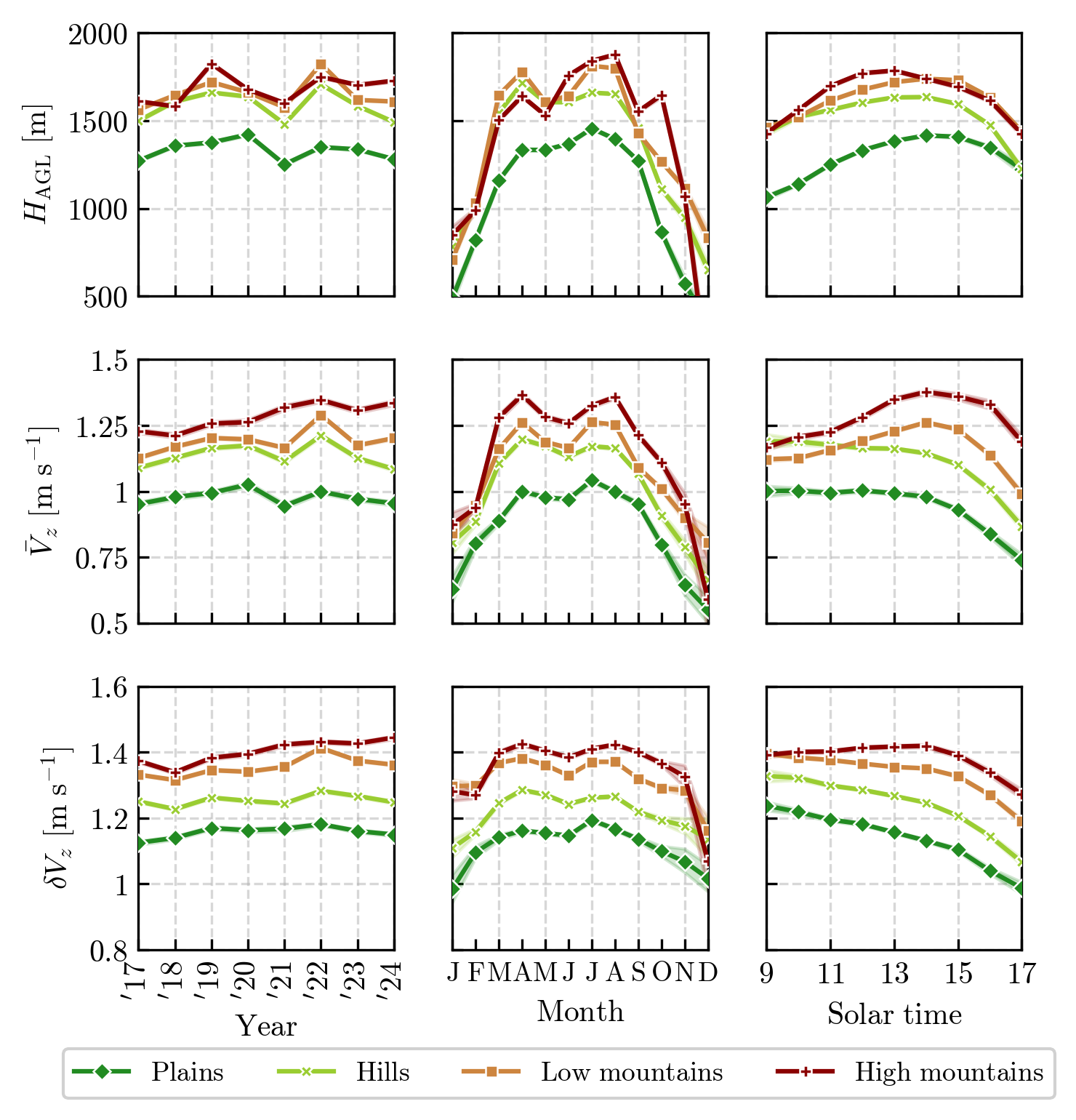}
\caption{Temporal evolution of the average climbing observables across three time scales.
Rows show thermal ceiling $\Hagl$ (top), thermal strength $\Vz$ (middle),
and vertical-velocity variability $\sigVz$ (bottom).
Columns show interannual evolution (left), seasonal or monthly evolution
(centre), and diurnal evolution (right, solar hour).
Lines are coloured by terrain class.}
\label{fig:stats}
\end{figure}

Figure~\ref{fig:stats} shows the temporal evolution of the three observables across three time scales. All observables exhibit moderate year-to-year variability without a strong overall secular trend, except for thermal strength and variability in High mountains, which shows a steady increase over the years. The exceptionally hot and dry  2022 summer in France stands out with elevated ceilings and thermal strength across all terrains. As expected, the seasonal cycle (middle column) reveals a strong modulation of the ceiling between winter ($\sim$500m~AGL) and summer ($\sim$1800m~AGL), driven by the seasonal evolution of surface heating. Thermal strength peaks in June--July and decays through autumn, while turbulence follows a similar but weaker pattern. Thermal strength peaks between 13--15h solar time (right column), consistent with the solar radiation cycle, with High mountains leading Plains by $\sim$1 hour due to earlier onset of convection at elevation.  The monotonic diurnal decrease in $\sigVz$ likely reflects the progressive organization of daytime convection. In the morning transition, newly developing thermals are shallow, intermittent, and highly variable, whereas later in the day the convective boundary layer deepens and the updrafts are more coherent and predictable. The terrain hierarchy High mountains $>$ Low mountains $>$ Hills $>$ Plains is remarkably robust across all three observables and all three time scales.

\subsection{Cell-hour aggregation}
\label{sec:aggregation}
The dataset of climbing segments is event-driven and spatially inhomogeneous: some cell-hour bins contain hundreds of climbing segments, while others contain none. To match the ERA5 spatio-temporal resolution ($0.25^\circ$, hourly) and to enable a direct correlation between climbing observables and reanalysis fields, we aggregate the segment-level data at the cell-hour level. For the ceiling, the aggregation~(\ref{eq:hagl}) already takes the maximum over segments within a cell-hour bin. 
For the velocity moments, consistent with the working hypothesis that thermal characteristics can be considered constant over one hour, we average the segment-level moments to obtain the cell-hour means:
{\begin{align}
    \Emvz &= \frac{1}{N_{c,d,h}} \sum_{i=1}^{N_{c,d,h}} \bar{V}_{z,i} \,,\label{eq:E_Vz}\\
    \Esvz &= \frac{1}{N_{c,d,h}} \sum_{i=1}^{N_{c,d,h}} \delta{V_z,i}\,, \label{eq:E_sVz}
\end{align}}
where $N_{c,d,h}$ is the number of climbing segments in cell $c$ on date $d$ at hour $h$. 

\subsection{Conditional splits}
\label{sec:splits}
To probe the physical predictors of thermal activity across different atmospheric regimes, we partition the dataset along four independent axes: season, time of day, cloud cover, and soil moisture state. All splits are applied \emph{within} each terrain class.\\

\paragraph*{Seasonal split.}
We define two seasons based on the annual cycle of solar insolation in metropolitan France:
\begin{equation}
    \text{Season} = \begin{cases}
        \text{Warm} & \text{month} \in \{3,4,5,6,7,8\} \\
        \text{Cold} & \text{month} \in \{9,10,11,12,1,2\}
    \end{cases}
    \label{eq:season}
\end{equation}
The warm season captures 85.6\% of all climbing segments, reflecting the well-known seasonality of thermal soaring. The cold season, while data-sparse, provides a physically contrasting regime with weaker surface heating.\\

\paragraph*{Diurnal split.}
We partition intra-day activity (solar hours 9--17h) into four blocks capturing distinct phases of thermal evolution:
\begin{equation}
    \text{Hour block} = \begin{cases}
        \text{All day}   & h \in [9, 17] \\
        \text{Morning}   & h \in [9, 11] \\
        \text{Midday}    & h \in [12, 14] \\
        \text{Afternoon} & h \in [15, 17]
    \end{cases}
    \label{eq:hours}
\end{equation}

\paragraph*{Cloud cover split.}
We separate clear-sky thermals from cloud-topped thermals using the ERA5 medium cloud cover fraction, namely $\mathtt{mcc}$:
\begin{equation}
    \text{Cloud cover state} = \begin{cases}
        \text{Clear}  & {\mathtt{mcc}} < 0.1 \\
        \text{Cloudy} & {\mathtt{mcc}} \geq 0.1
    \end{cases},
    \label{eq:cloud}
\end{equation}
Although it may seem counterintuitive, we use the cloud fraction in the mid-tropospheric layer ($\sim$2--6km ASL) rather than low cloud cover ({$\mathtt{lcc}$}) because most climbing segments are in mountainous terrain, where a cloud formed at 2500m~ASL over a 1500m mountain is a mid-level cloud by the ERA5 layer definition. The threshold $\tau=0.1$ (10\% cloud coverage) matches the standard operational definition of clear vs cloudy skies~\footnote{\small\url{https://forecast.weather.gov/glossary.php?word=sky\%20condition}}.\\

\paragraph*{Soil-moisture split.}
Thermal generation is strongly influenced by the surface energy budget, which in turn depends on soil moisture. However, absolute soil moisture ($\theta_{\ell1}$, the ERA5 volumetric soil water in layer~1, {$\mathtt{swvl1}$}) is not directly comparable across terrain and soil types. We instead use the Soil Moisture Index (SMI)~\cite{Betts2004, Johannsen2020}:
\begin{equation}
    \mathrm{SMI} = \frac{\theta_{\ell1} - \theta_{\mathrm{wp}}}{\theta_{\mathrm{fc}} - \theta_{\mathrm{wp}}}\,,
    \label{eq:smi}
\end{equation}
where $\theta_{\mathrm{fc}}$ and $\theta_{\mathrm{wp}}$ are the soil-type-dependent field capacity and wilting point respectively, indexed by the ERA5 soil-type field ($\mathtt{slt}$) and derived  from 
the H-TESSEL land-surface scheme~\cite{Balsamo2009}. SMI is a dimensionless quantity, where SMI = 0 corresponds to fully wilted soil (no available water for plants), SMI = 1 corresponds to soil at field capacity, and SMI $> 1$ indicates soil moisture above field capacity (waterlogged conditions). The soil split is given by:
\begin{equation}
    \text{Soil state} = \begin{cases}
        \text{Dry} & \mathrm{SMI} < 0.5 \\
        \text{Wet} & \mathrm{SMI} \geq 0.5
    \end{cases},
    \label{eq:soil}
\end{equation}
where the $\mathrm{SMI} = 0.5$ threshold  (soil at half of plant-available water capacity) is the standard operational cutoff between dry and wet conditions.

The four splits can be combined multiplicatively to define finer regimes such as ``Warm midday clear-sky dry-soil'' (denoted $\mathtt{Warm\_Midday\_Cl\_Dry}$ in the following).  The next section characterizes the physical variability captured by these conditional splits.

\subsection{Summary statistics of conditional splits}
\label{sec:stats}
For each combination of terrain, season, hour block, cloud cover state, and soil-moisture state, we compute the number of cell-hour observations $n$ and the sample mean of each thermal observable. This descriptive step serves three purposes: (i) validate that the splits capture genuine physical variability rather than statistical noise; (ii) reveal which combinations carry the largest signal and are therefore most informative; and (iii) motivate the choice of a reference regime for the driver analysis in Section~\ref{sec:results}.\\

\paragraph*{Terrain, season and diurnal effects.}
Computing the averages of the three climbing metrics for the four terrain classes, two seasons, and four hour blocks (cloud and soil combined) revealed three robust patterns.

\begin{enumerate}
    \item \emph{Terrain hierarchy.} Across all splits, the ordering \emph{High mountains} $>$ \emph{Low mountains} $>$ \emph{Hills} $>$ \emph{Plains} is preserved for the three observables. In the Warm season, the thermal ceiling ranges from $\bar{H}_{\mathrm{AGL}} = 1613$m in High mountains to 1313m in Plains ($\sim 23\%$ range), the mean climb rate from 1.35 to 1.01ms$^{-1}$ ($\sim 34\%$ range), and the turbulence from 1.40 to 1.16ms$^{-1}$ ($\sim 21\%$ range).

    \item \emph{Seasonal contrast.} Comparing Warm to Cold within Low mountains, the summary statistics give $\bar{H}_{\mathrm{AGL}} = 1529$m vs 1199m ($-22\%$), $\bar{V}_z = 1.20$~vs 1.01ms$^{-1}$ ($-16\%$), and $\delta{V_z} = 1.33$~vs 1.27ms$^{-1}$ ($-5\%$). The ceiling is much lower in the Cold season. The climb rate is moderately weaker, whereas turbulence intensity is nearly season-invariant.

    \item \emph{Diurnal cycle.} Within the Warm season, all three observables show a midday peak. For Low mountains, midday (12--14h) gives $\bar{H}_{\mathrm{AGL}} = 1598$m compared to 1443m in the morning ($+11\%$) and 1488m in the afternoon ($+7\%$). The mean climb rate follows a similar pattern with slightly weaker amplitude. $\delta{V_z}$ shows a different diurnal signature: it is highest in the morning (1.38ms$^{-1}$) and decays monotonically through the day, reaching 1.25ms$^{-1}$ in the afternoon, consistent with the progressive organization of daytime convection discussed above.
\end{enumerate}

\paragraph*{Cloud and soil-moisture effects at midday.}
Here we explore the effect of the cloud state and the soil moisture at Warm midday across the four terrains.

\begin{enumerate}
    \item \emph{Soil moisture effect.} The Dry-vs-Wet effect on $\bar{H}_{\mathrm{AGL}}$ is large and terrain-consistent: for Low mountains, dry-soil conditions give $\bar{H}_{\mathrm{AGL}} = 1883$m vs 1492m for wet soil, a $+26\%$ variation. The effect ranges from $+12\%$ in Plains to $+26\%$ in Low mountains, is preserved across all terrains, and applies with similar amplitude to the climb rate ($\bar{V}_z = 1.42$~vs 1.17ms$^{-1}$ in Low mountains, $+21\%$).

    \item \emph{Cloud state effect.} The Clear-vs-Cloudy effect on $\bar{H}_{\mathrm{AGL}}$ is small and mixed in sign: for Low mountains, clear conditions give a marginally higher ceiling (1626~vs 1565m, cloudy conditions give a $3.7\%$ lower ceiling), while for Plains the sign flips (Cloudy 1403m~vs Clear 1321m, $+6\%$ for cloudy). High mountains show a similar cloudy enhancement ($+2.5\%$), and Hills are essentially unaffected.

    \item \emph{Additivity of cloud and soil effects.} Comparing the marginals splits with their intersection, the two effects are approximately additive: for Low mountains, the Clear-skies split deviates from the baseline unconditional split by $+28$m, the Dry-soil by $+285$m, and the Clear+Dry intersection by $+322$m ($\approx 28+285 = 313$m, small positive interaction).
\end{enumerate}

Tables \ref{tab:stats_seasonal_diurnal} and \ref{tab:stats_cloud_soil} in  Appendix~\ref{sec:stats_tab} report the means of the climbing metrics across terrains and conditional splits.

\section{ERA5-based driver identification}
\label{sec:results}
\subsection{Motivation}
The summary statistics presented above reveal three main features of the flight logs dataset. First, the terrain hierarchy High mountains $>$ Low mountains $>$ Hills $>$ Plains is a robust feature across all combinations and reflects the effect of orography on thermal convection development.
Second, the seasonal and diurnal cycles produce large modulations of the ceiling and climb rate ($10$--$25\%$) but only modest changes in turbulence intensity ($<10\%$). Third, at fixed hour and season, soil moisture is the dominant conditional split, with the Dry-vs-Wet effect on the ceiling reaching $+26\%$ in Low mountains;  cloud state alone contributes a much weaker, terrain-dependent modulation.

Together, these findings guide the design of a correlation and regression analysis. The midday block (12--14h solar) combines near-peak values of all three observables and concentrates the bulk of the data: it accounts for $\sim$$55\%$ of all cell-hour observations in the warm season and $\sim$$67\%$ in the cold season. Because cloud and soil state are shown here to be secondary compared to terrain and season, and because further splitting by cloud$\times$soil rapidly depletes the per-cell sample, we retain the four terrains and two seasons at midday ($\mathtt{Warm\_Midday\_All}$ and $\mathtt{Cold\_Midday\_All}$) as the reference regimes for our analysis. 

Below, we use the cell-hour aggregated climbing observables to identify which ERA5 atmospheric variables drive their behavior. We first reduce the 206 raw ERA5 variables to a physically interpretable set of 120 predictors, then establish the well-known {lifting condensation level (LCL)} scaling as a first-order benchmark, and finally use an orthogonal framework to identify the strongest independent secondary predictors of each observable after removing the contribution of dew point depression $T-T_d$ ($= \mathtt{t2m}-\mathtt{d2m}$ with the ERA5 notation) from the correlation signal.

\subsection{ERA5 variable selection}
\label{sec:era5vars}
The full ERA5 extraction of 206 variables includes many quantities that are physically redundant or unsuitable for our purposes. We thus apply a four-stage physical filter as follows. First, we remove variables specific to ocean or glacier surfaces, since our analysis focuses on land-based thermal convection over metropolitan France. We then exclude five snow-related variables, including snowfall, snow water equivalent, and related fluxes, because they are zero for the vast majority of climbing segments, which typically occur under warm conditions or in mountains during snow-free periods. Next, we discard fourteen zero-inflated variables whose fraction of exactly zero values exceeds 50\% of observations; these include large-scale precipitation, convective precipitation rate, several runoff and drainage variables, and CAPE. Finally, we remove fourteen variables on physical or logical grounds: dew point temperature $\mathtt{d2m}$, whose information is fully carried by our benchmark predictor $(\ttd)$ once $\mathtt{t2m}$ is retained; medium cloud cover {$\mathtt{mcc}$}, which is used to define the cloud split; soil moisture at layer~1, which is used to define the SMI split; total cloud cover {$\mathtt{tcc}$}, which is a composite of {$\mathtt{lcc}$}, {$\mathtt{mcc}$}, and {$\mathtt{hcc}$} and therefore redundant; the instantaneous versions of several fluxes ($\mathtt{ishf}$, $\mathtt{ie}$, $\mathtt{iews}$, $\mathtt{inss}$), which are numerically identical to their time-averaged $\mathtt{avg\_}$-prefixed counterparts; and the log-transformed roughness $\mathtt{flsr}$, which is redundant with $\mathtt{fsr}$.

One is left with 120 predictors which span all relevant physical variable families (see Table~\ref{tab:era5_predictors} in Appendix~\ref{ap:vars}): near-surface temperature ($\mathtt{t2m}$, $\mathtt{skt}$, $\mathtt{mn2t}$, $\mathtt{mx2t}$, $\mathtt{stl1}$--$\mathtt{stl4}$), moisture ($\mathtt{t2m}-\mathtt{d2m}$ and column-integrated water content, including water vapor and cloud water/ice, e.g., $\mathtt{tcwv}$), radiation (shortwave and longwave, both surface and top-of-atmosphere, all-sky and clear-sky), turbulent fluxes ($\mathtt{avg\_ishf}$, $\mathtt{avg\_ie}$, $\mathtt{avg\_iews}$, $\mathtt{avg\_inss}$), soil moisture at three remaining layers ($\mathtt{swvl2}$--$\mathtt{swvl4}$), boundary-layer variables ($\mathtt{blh}$, cloud base height), surface characteristics (albedos $\mathtt{alnip}$/$\mathtt{alnid}$/$\mathtt{aluvd}$/$\mathtt{aluvp}$, roughness $\mathtt{fsr}$, leaf-area indices $\mathtt{lai\_lv}$/$\mathtt{lai\_hv}$), wind vectors and derived speeds at 10m and 100m ($\mathtt{u10}$, $\mathtt{v10}$, $\mathtt{uv10}$, $\mathtt{u100}$, $\mathtt{v100}$, $\mathtt{uv100}$), and vertically integrated atmospheric quantities.
The full list of 120 predictors with physical descriptions and units is provided in Appendix~\ref{ap:vars}.

\subsection{The LCL scaling as a first-order benchmark}
\label{sec:lcl}
For a cloud-topped or wet thermal, the classical lifting condensation level (LCL) theory predicts a top height proportional to the dew point depression $\ttd$, with a slope of $\sim$$125$m/$^\circ$C for dry-adiabatic ascent of a well-mixed parcel~\cite{Bolton1980,Lawrence2005,Romps2017}. This scaling does not apply to purely dry thermals, whose ceiling is instead set by the boundary-layer depth and the capping inversion above it~\cite{Pergaud2009}. Beyond this scaling, atmospheric convection involves many other physical predictors (boundary-layer depth, surface energy fluxes, radiation, soil moisture etc.)  that are not captured by a single-variable relationship. Nonetheless, the LCL relation $\Hagl \propto \ttd$ provides a natural first-order benchmark to test against the paragliding-derived ceiling: it is thermodynamically grounded, involves only two ERA5 variables {($\mathtt{t2m}$ and $\mathtt{d2m}$)}, and gives a concrete  prediction for the slope.

We test this scaling with an origin-constrained ordinary least squares (OLS) linear regression for the
ceiling,
\begin{equation}\label{eq:lcl_benchmark_H}
    \Hagl = \beta_1 \cdot (T - T_d)\,, \qquad \gamma \equiv 0\,,
\end{equation}
and standard OLS linear regression with intercept for the two velocity moments,
\begin{equation}\label{eq:lcl_benchmark_Vz}
    Y = \beta_1 \cdot (T - T_d) + \gamma\,, \qquad
    Y \in \{\Emvz, \Esvz\}\,,
\end{equation}
applied to all cell-hour observations within a given (terrain, season) combination at midday.
To assess how well each regression captures the observed relationship, we complement the OLS slope $\beta_1$ with the Spearman rank correlation $\rho_S(Y, T-T_d)$ between the observable and the dew point depression, which quantifies the strength of any monotonic association\footnote{Note that Pearson linear correlations $r(Y, T-T_d)$ agree with $\rho_S$ to within $\pm 0.03$ across all combinations reported below.}.

\begin{figure}[t!]
 \centering
\includegraphics[width=\columnwidth]{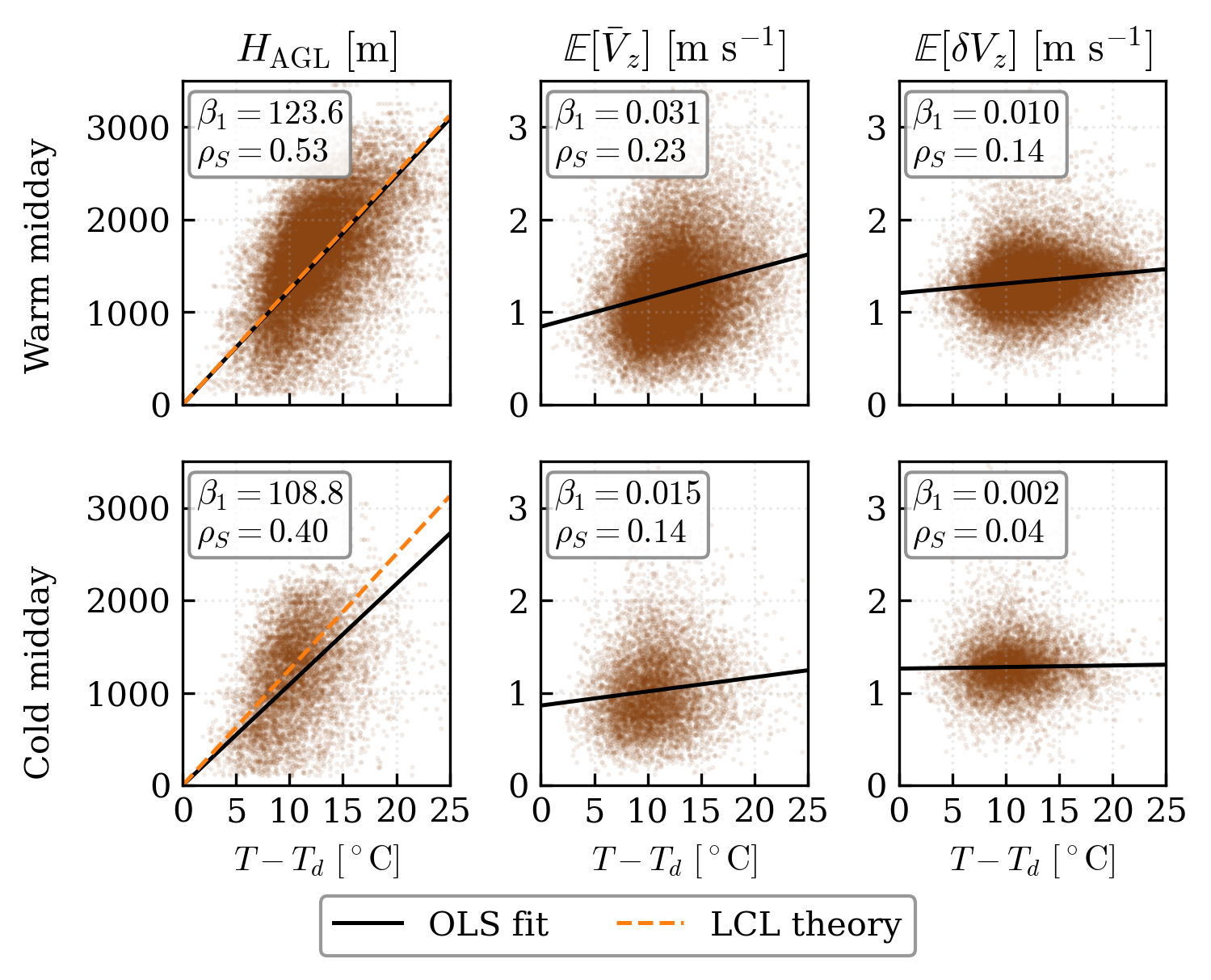}
\caption{LCL benchmark scatter for the Low mountains terrain at midday (12--14h solar). Rows: Warm season (top, $n = 29{,}023$ cell-hours) and Cold season (bottom, $n = 8{,}217$). Columns: thermal ceiling $\Hagl$ (left), mean climb rate $\Emvz$ (centre), and climb turbulence $\Esvz$ (right), each plotted against the dew point depression $T{-}T_d$. Each brown dot is one cell-hour observation. The black solid line is the OLS fit (origin-constrained for $\Hagl$, Eq.~\ref{eq:lcl_benchmark_H}; with intercept for the velocity moments, Eq.~\ref{eq:lcl_benchmark_Vz}). In the $\Hagl$ panels, the orange dashed line shows the theoretical LCL slope of $125$m/$^\circ$C. Each panel indicates the corresponding fitted slope $\beta_1$ (m/$^\circ$C for $\Hagl$; ms$^{-1}$/$^\circ$C for the velocity moments) and Spearman correlation $\rho_S(Y, T{-}T_d)$.}
\label{fig:lcl_scatter}
\end{figure}

Figure~\ref{fig:lcl_scatter} illustrates the LCL benchmark for the Low mountains terrain, chosen as the reference case: it contains the largest number of climbing segments among the four terrains ($\sim$$37\,000$ cell-hours over the two seasons) and covers a broad range of altitudes, soil types, and vegetation. Three visual features are worth underlining. First, in the Warm season (top-left panel), the OLS  fit is nearly indistinguishable from the theoretical LCL line: the empirical slope $\beta_1 = 123.6$m/$^\circ$C matches the theoretical value of $125$m/$^\circ$C to within $1\%$, providing a physically grounded calibration of the paragliding-derived ceiling. 
Second, in the Cold season (bottom-left panel), the two lines visibly diverge: the empirical slope drops to $108.8$m/$^\circ$C, $\sim$$13\%$ below theory. This likely reflects a combination of physical effects, including a limitation of our free-flight-based observable. In the Cold season, thermals are weaker, so the paraglider’s sink rate may become comparable to the atmospheric updraft velocity, making it harder for pilots to reach the true thermal ceiling. As a result, the $\Hagl$ estimator acts as a more severe lower bound on the actual ceiling in the Cold season than in the Warm one.
Third, for the velocity moments (center and right columns) the point clouds show only a mild tilt with $\ttd$ in the Warm season and are essentially flat in the Cold season, consistent with the small $\rho_S$ values in the legends. 
Table~\ref{tab:lcl_benchmark} extends the analysis to the four terrains, two seasons and three targets where three patterns stand out.

\begin{table}[t!]
\centering
\caption{LCL benchmark regressions at midday (12--14h solar) for the four
terrains and two seasons. For each target $Y$ we report the slope
$\beta_1$ (m/$^\circ$C for $\Hagl$; [m\,s$^{-1}$]/$^\circ$C for the velocity
moments) and the Spearman correlation $\rho_S$ between $Y$ and
$T{-}T_d$. }
\label{tab:lcl_benchmark}
\renewcommand{\arraystretch}{1.05}
\setlength{\tabcolsep}{2.5pt}
\begin{tabular}{llr r r r r r r}
\toprule
 & & & \multicolumn{2}{c}{$\Hagl$}
     & \multicolumn{2}{c}{$\Emvz$}
     & \multicolumn{2}{c}{$\Esvz$} \\
\cmidrule(lr){4-5}\cmidrule(lr){6-7}\cmidrule(lr){8-9}
Terrain & Season & $n$ & $\beta_1$ & $\rho_S$
                     & $\beta_1$ & $\rho_S$
                     & $\beta_1$ & $\rho_S$ \\
\midrule
High m. & Warm & 8{,}648  & 133.3 & 0.47 & 0.023 & 0.17 & 0.008 & 0.12 \\
               & Cold & 3{,}110  & 117.4 & 0.27 & 0.017 & 0.12 & 0.007 & 0.09 \\
\midrule
Low m.  & Warm & 29{,}023 & 123.6 & 0.53 & 0.031 & 0.23 & 0.010 & 0.14 \\
               & Cold & 8{,}217  & 108.8 & 0.40 & 0.015 & 0.14 & 0.002 & 0.04 \\
\midrule
Hills          & Warm & 24{,}211 & 106.4 & 0.46 & 0.016 & 0.15 & 0.007 & 0.11 \\
               & Cold & 5{,}050  &  98.3 & 0.44 & 0.018 & 0.19 & 0.008 & 0.13 \\
\midrule
Plains         & Warm & 14{,}459 & 100.9 & 0.42 & 0.012 & 0.11 & 0.004 & 0.06 \\
               & Cold & 2{,}646  &  91.1 & 0.54 & 0.023 & 0.28 & 0.009 & 0.19 \\
\bottomrule
\end{tabular}
\end{table}

\begin{enumerate}
    \item In the Warm season, the empirical slope for $\Hagl$ decreases monotonically from $133$m/$^\circ$C in High mountains through $124$m/$^\circ$C in Low mountains and $106$m/$^\circ$C in Hills down to $101$m/$^\circ$C in Plains. The two mountainous terrains bracket the textbook value of $\sim$$125$m/$^\circ$C, while the flatter terrains fall $\sim$$15$--$20\%$ below it.

    \item Within each (terrain, season) configuration, the correlation strength follows $\rho_S(\Hagl,\ttd) > \rho_S(\Emvz,\ttd) > \rho_S(\Esvz,\ttd)$ without exception across all eight configurations. For $\Hagl$ in the Warm season, $\rho_S$ lies in the range $0.42$--$0.53$, indicating a moderate but genuine link. For the mean climb rate $\Emvz$, $\rho_S$ drops to $0.11$--$0.23$, and for turbulence $\Esvz$ to $0.06$--$0.14$. The LCL scaling therefore correctly identifies height (altitude) physics but says little about the dynamics of buoyancy generation or small-scale mixing within the thermals, as expected. This gap motivates the orthogonal regression analysis below, which asks which \emph{other} atmospheric variables in the ERA5 database carry independent information about each observable.

    \item Moving from Warm to Cold at fixed terrain, $\rho_S(\Hagl,\ttd)$ drops from $0.47$ to $0.27$ in High mountains and from $0.53$ to $0.40$ in Low mountains, while it remains essentially flat in Hills ($0.46 \to 0.44$) and \emph{increases} from $0.42$ to $0.54$ in Plains. The empirical slope $\beta_1$ decreases in the mountains ($133 \to 117$, $124 \to 109$m/$^\circ$C) but drops much less in Plains ($101 \to 91$m/$^\circ$C). Interpreting the Plains Cold value requires caution, the sample is limited to $n = 2{,}646$ cell-hour observations, so sampling variability is non-negligible. Furthermore, the ceiling under-reach effect discussed above compounds this uncertainty, since weak thermals in Cold-season Plains conditions may be especially difficult to top out. Still, the qualitative signal is robust: in the Cold season, mountainous thermals decouple from $\ttd$, while ceiling in the plains remains partly $\ttd$-driven.
\end{enumerate}

\subsection{Orthogonalized rank-feature relevance}
\label{sec:orfr}

\begin{figure*}[p]
\centering
\includegraphics[width=0.49\textwidth]{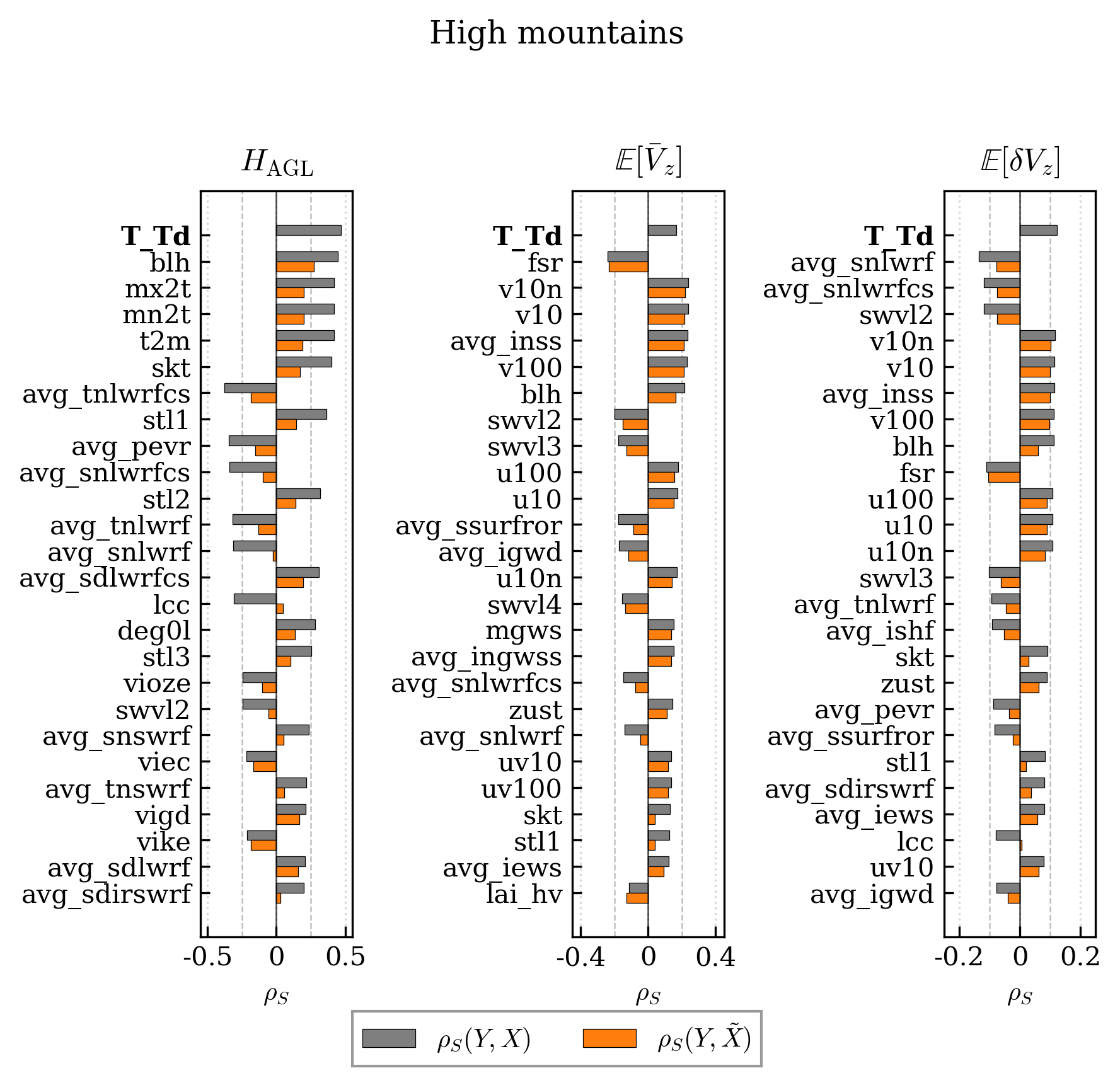}\hfill
\includegraphics[width=0.49\textwidth]{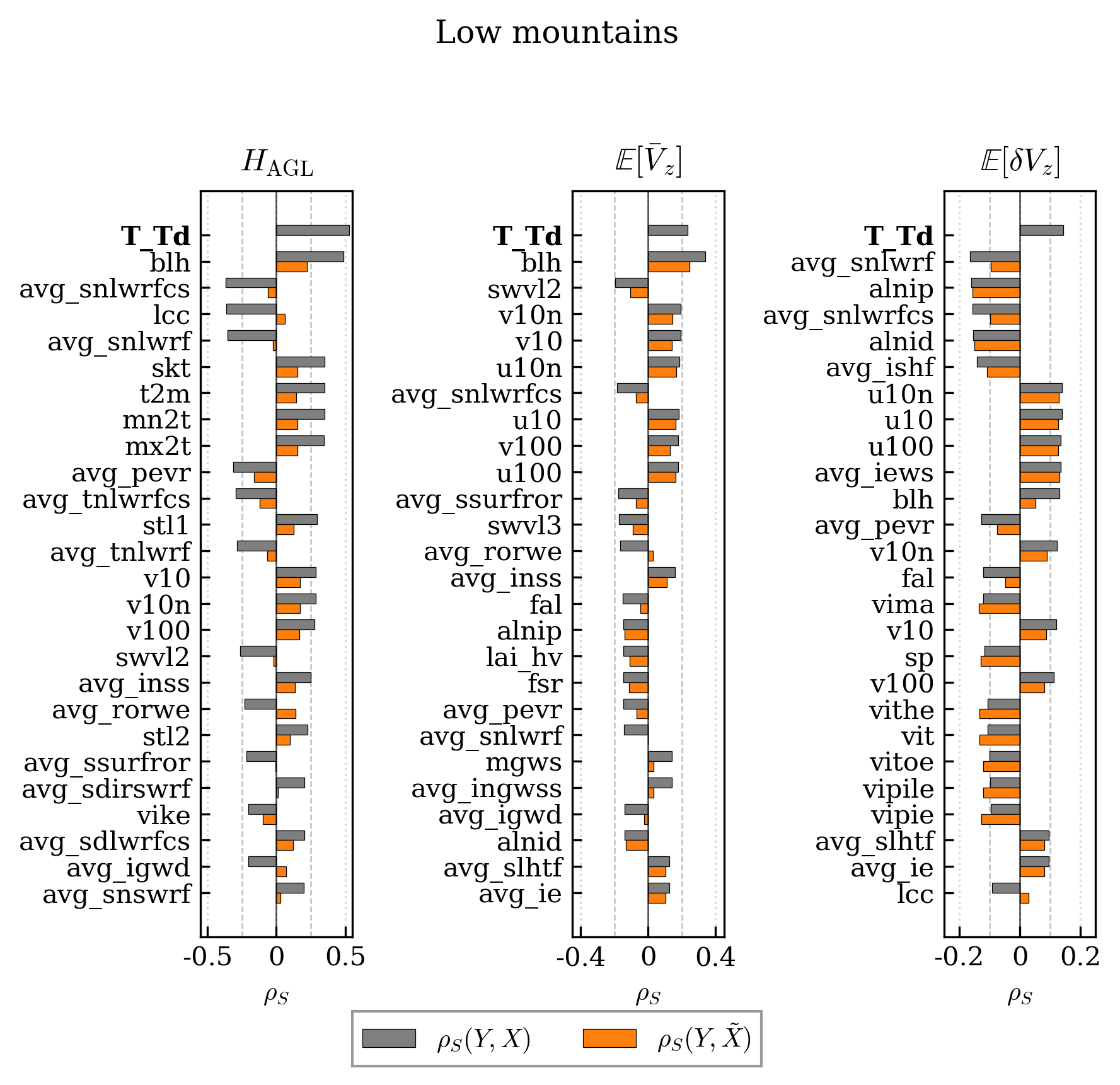}\\[0.5em]
\includegraphics[width=0.49\textwidth]{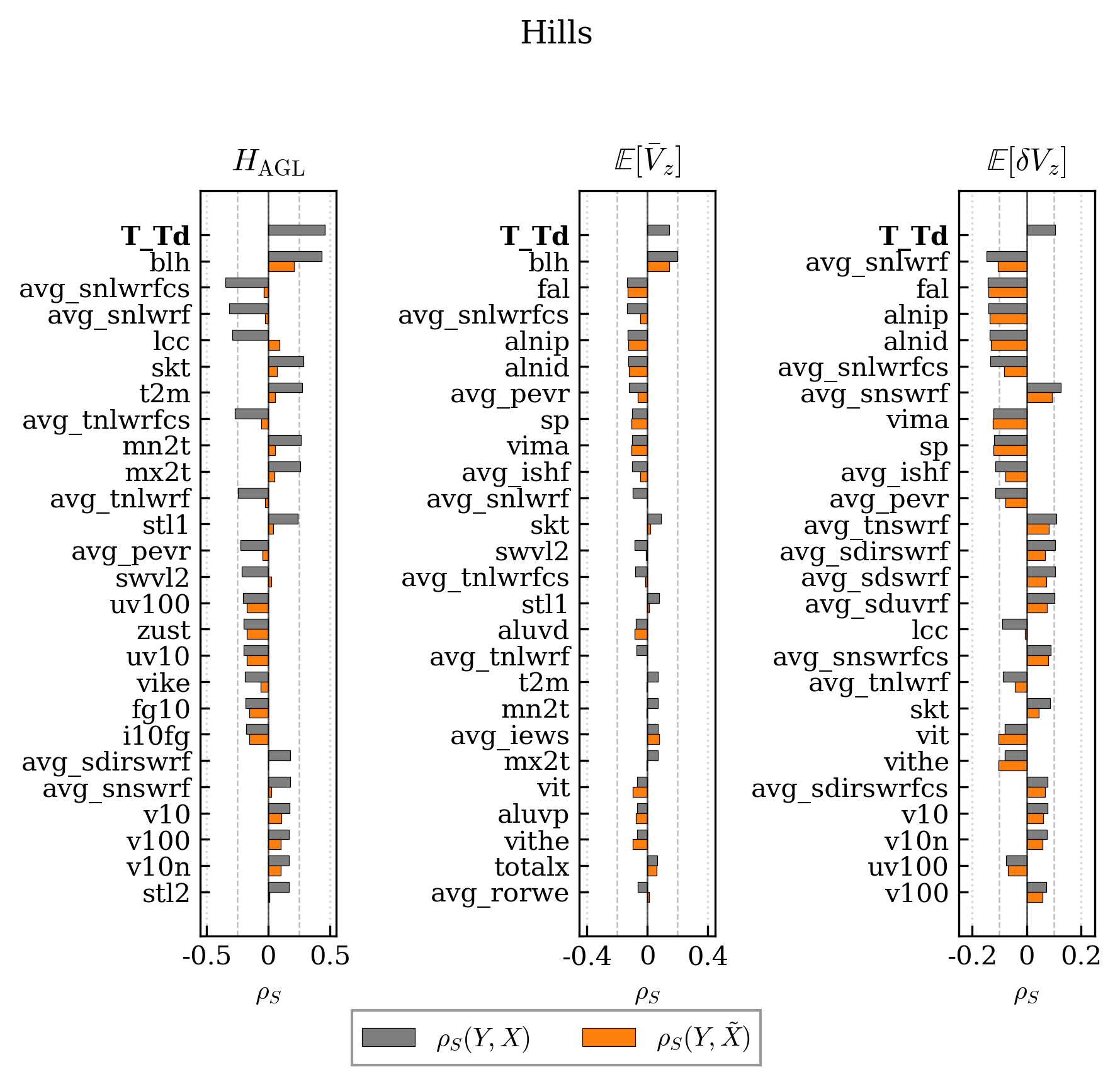}\hfill
\includegraphics[width=0.49\textwidth]{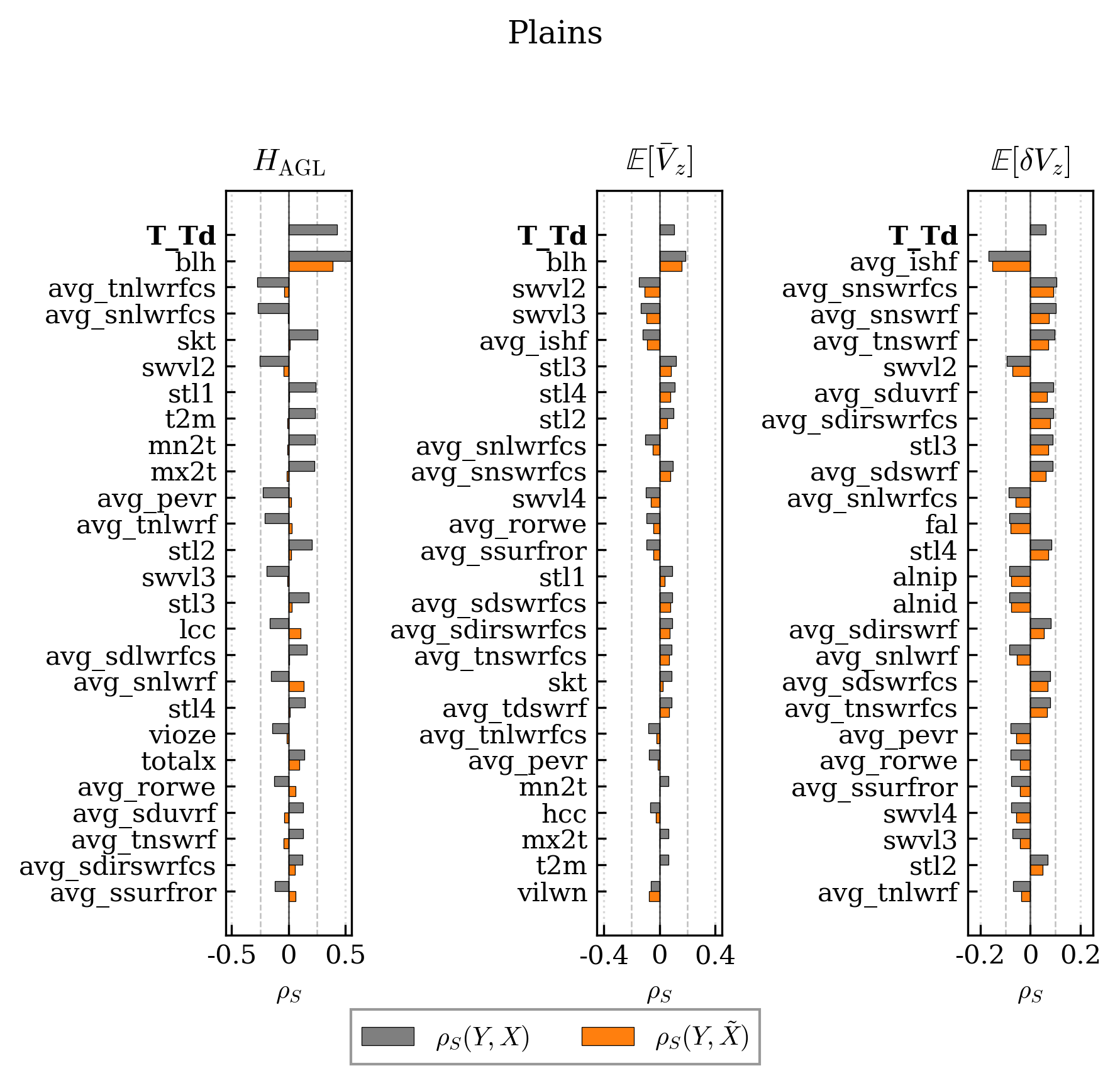}
\caption{ORFR bar plots at Warm midday (12--14h solar) for the four terrains, in reading order: High mountains (top-left, $n = 8{,}648$), Low mountains (top-right, $n = 29{,}023$), Hills (bottom-left, $n = 24{,}211$), and Plains (bottom-right, $n = 14{,}459$). Each panel ranks the top-25 ERA5 predictors of $\Hagl$ (left), $\Emvz$ (centre), and $\Esvz$ (right) by $|\rho_S(Y, X_j)|$. Grey bars indicate raw Spearman correlation $\rho_S(Y, X)$; orange bars are correlation after $\ttd$-orthogonalization, $\rho_S(Y, \tilde{X})$. The top grey bar ($\mathtt{T\_Td}$, in bold) is the reference correlation.}
\label{fig:orfr_warm}
\end{figure*}

\begin{figure*}[p]
\centering
\includegraphics[width=0.49\textwidth]{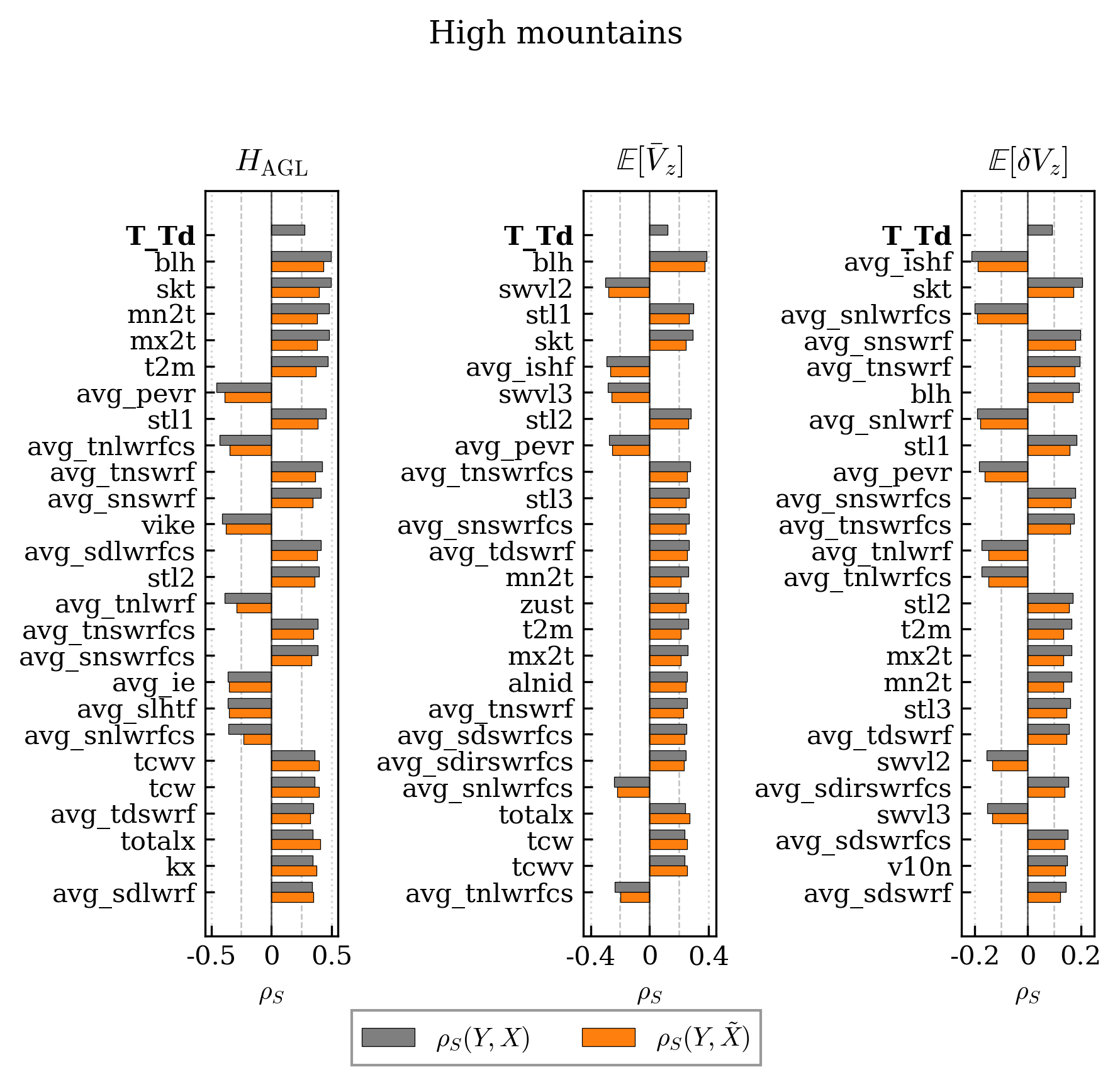}\hfill
\includegraphics[width=0.49\textwidth]{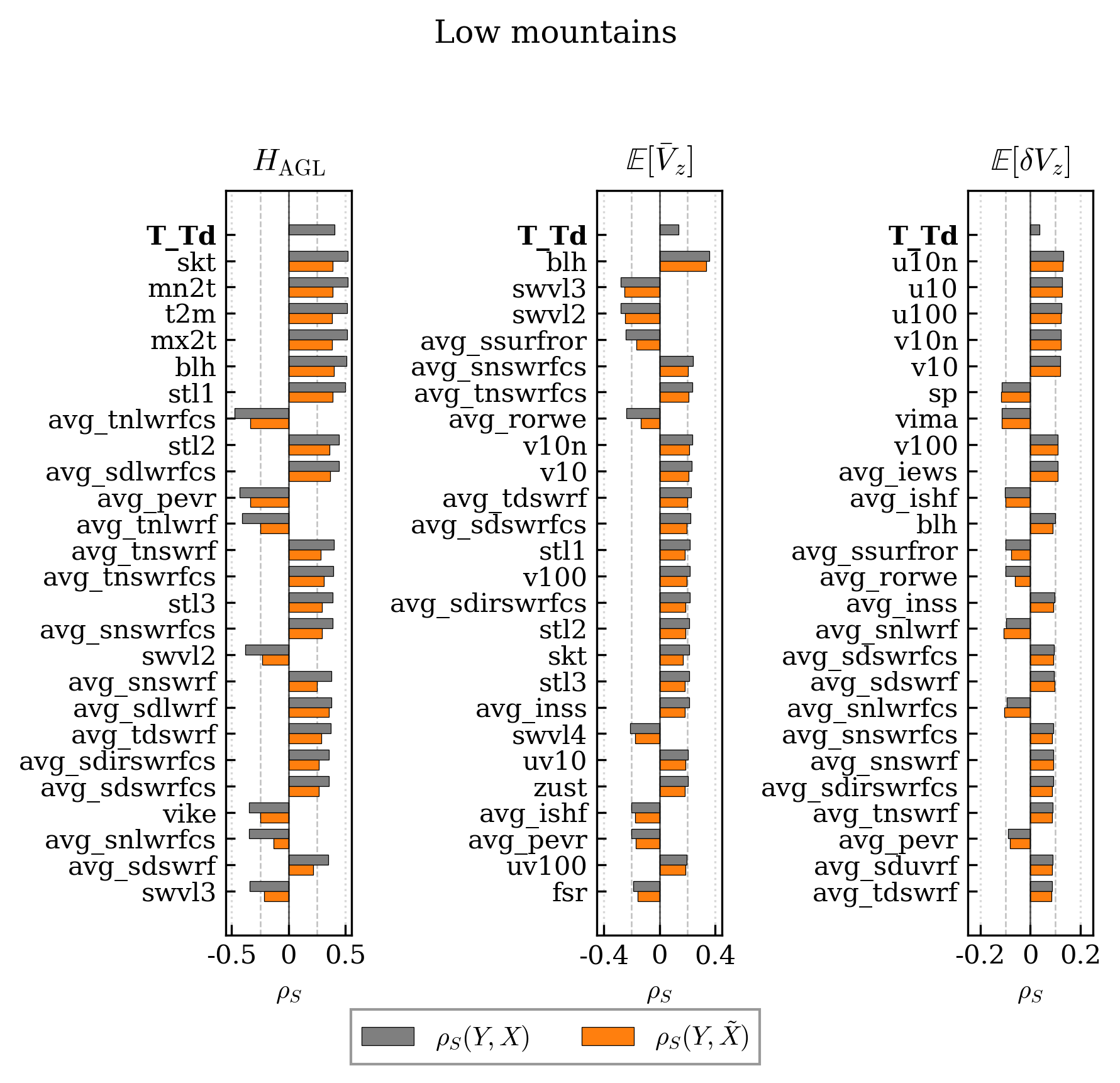}\\[0.5em]
\includegraphics[width=0.49\textwidth]{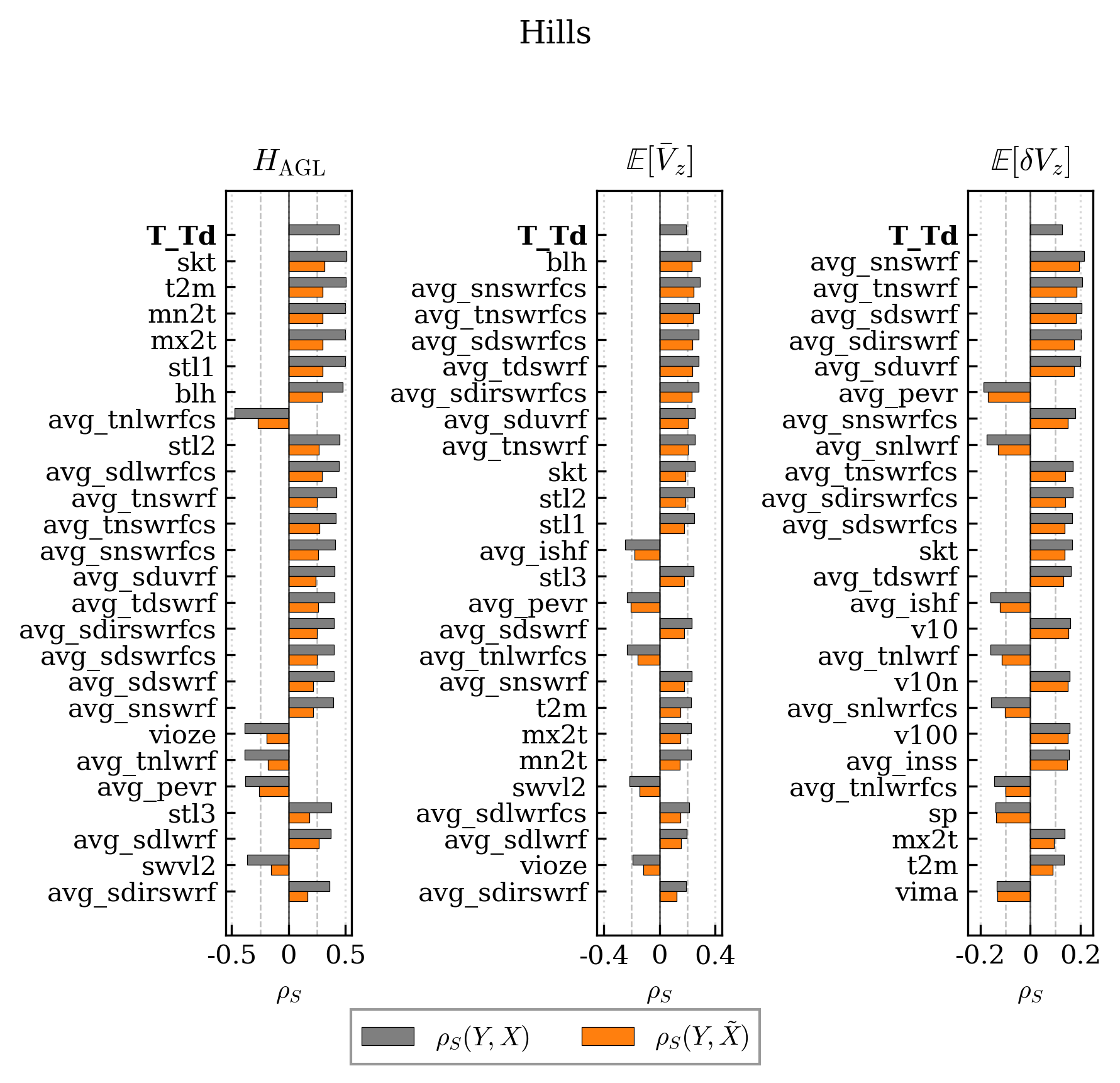}\hfill
\includegraphics[width=0.49\textwidth]{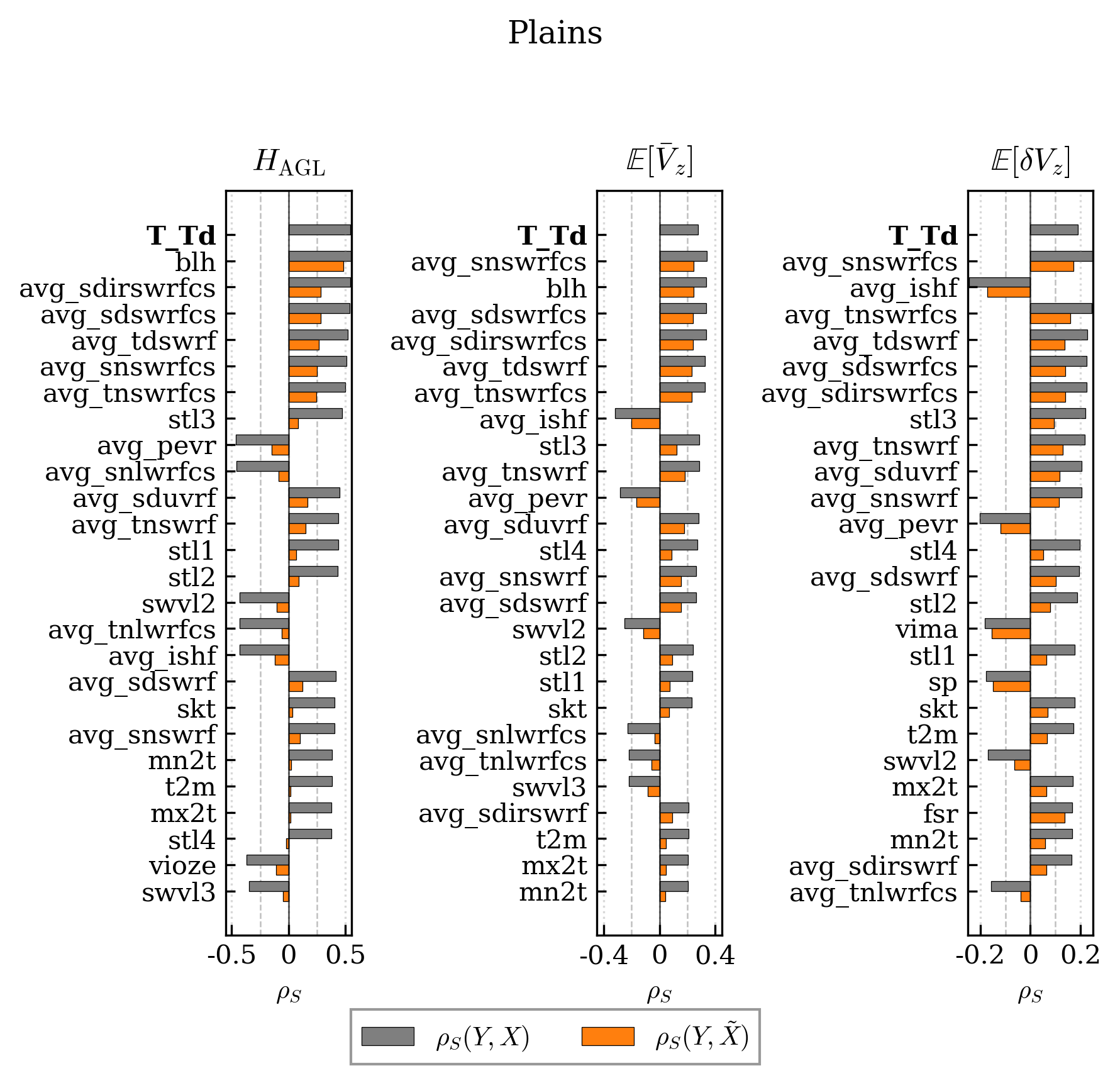}
\caption{ORFR bar plots at Cold midday (12--14h solar) for the four terrains, in the same reading order: High mountains (top-left, $n = 3{,}110$), Low mountains (top-right, $n = 8{,}217$), Hills (bottom-left, $n = 5{,}050$), and Plains (bottom-right, $n = 2{,}646$). Panels and bars follow the convention of Fig.~\ref{fig:orfr_warm}.}
\label{fig:orfr_cold}
\end{figure*}

To identify variables that carry information about the thermal observables \emph{beyond} the LCL scaling, we use an orthogonal regression procedure called orthogonalized rank-feature relevance (ORFR). The core idea is simple: for each ERA5 predictor, we first remove the component of variability that can be explained by the dew point depression (the LCL reference), then rank all predictors by their correlation with the observable after this removal. Variables that retain strong correlation after this decorrelation step carry genuinely independent information about thermal properties. Formally, let $Y$ denote a thermal observable {$(\Hagl,\,\Emvz,\,\Esvz)$} and $Z \equiv \ttd$ the reference predictor. For each candidate variable $X_j$ ($j = 1, \ldots, 119$), we first remove its linear covariance with $Z$ by fitting an auxiliary regression:
\begin{equation}
    X_j = \alpha_{\mathrm{aux},j} + \beta_{\mathrm{aux},j} \cdot Z\,, \
    \tilde{X}_j = X_j - \alpha_{\mathrm{aux},j} - \beta_{\mathrm{aux},j} \cdot Z
    \label{eq:aux_B}
\end{equation}
where $\alpha_{\mathrm{aux}}$ and $\beta_{\mathrm{aux}}$ are estimated by standard OLS. This satisfies the equations: $\mathrm{cov}(Z, \tilde{X}_j) = 0$ and $\mathrm{mean}(\tilde{X}_j) =0$.

We then compute the Spearman rank correlation $\rho_S(Y, \tilde{X}_j)$  between the thermal observable and the residual, and rank all 119 variables by the absolute value of the correlation. Note that, as a robustness check, we also performed the analysis using Pearson linear correlation $r(Y, \tilde{X}_j)$ and found consistent results. The top-ranked variables are those that carry the strongest \emph{independent} information about $Y$ beyond what $\ttd$ already provides.

\paragraph*{Univariate regression on residuals.}
For each $(Y, X_j)$ pair we fit a univariate regression on the residual:
\begin{equation}
    Y = \beta'_j \cdot \tilde{X}_j + \alpha'_j\,,
    \label{eq:univariate}
\end{equation}
where $\beta'_j$ has units of $[Y]/[X_j]$ and represents the physical sensitivity of $Y$ to variations in $X_j$ beyond the $\ttd$-explained component. Because this method guarantees $\mathrm{mean}(\tilde{X}_j) = 0$, the intercept simplifies to $\alpha'_j = \bar{Y}$ for every variable $j$: the constant term is fixed at the sample mean and only the slope $\beta'_j$ varies across candidates. This gives the interpretable decomposition
\begin{equation}
    Y = \bar{Y} + \beta'_j \cdot \tilde{X}_j\,,
    \label{eq:decomposition}
\end{equation}
which reads as: the expected value of $Y$ equals its sample mean, modulated by anomalies of $X_j$ that are not attributable to the LCL scaling.

Figures~\ref{fig:orfr_warm} and \ref{fig:orfr_cold} display, for each of the four terrains and the two midday reference seasons, the 25 variables with the largest $|\rho_S(Y, X_j)|$ (i.e., ranked by the raw Spearman correlation, before orthogonalization) in each of the three targets $\Hagl$, $\Emvz$, and $\Esvz$, ranked in descending order from top to bottom. The grey bar shows the raw Spearman correlation $\rho_S(Y, X_j)$; the orange bar shows the correlation after orthogonalization, $\rho_S(Y, \tilde{X}_j)$. The reference correlation with $\ttd$ itself is drawn as the top grey bar (variable label $\mathtt{T\_Td}$, bold) and has no orange counterpart by construction. In what follows we describe the rankings in terms of both the raw Spearman correlation $\rho_S(Y, X_j)$ and its \emph{retention} after decorrelation, defined as $|\rho_S(Y, \tilde{X}_j)|\,/\,|\rho_S(Y, X_j)|$\footnote{Because the retention is defined as a ratio, it can become unstable when the raw correlation $|\rho_S(Y, X_j)|$ is small. To verify the robustness of our rankings, we recomputed the retention using two alternative definitions: the absolute difference $|\rho_S(Y, X_j)| - |\rho_S(Y, \tilde{X}_j)|$, and its normalized version $\big(|\rho_S(Y, X_j)| - |\rho_S(Y, \tilde{X}_j)|\big) / \max\big(|\rho_S(Y, X_j)|, |\rho_S(Y, \tilde{X}_j)|\big)$. The top-25 predictor set is essentially preserved.}. A retention close to unity indicates that the variable carries information about $Y$ that is essentially independent of $\ttd$; a retention close to zero indicates that the variable is largely a proxy for $\ttd$ at the cell-hour scale.

\paragraph*{Warm midday.}
Figure~\ref{fig:orfr_warm} shows the ORFR results at Warm midday. Across all four terrains, the boundary-layer height $\mathtt{blh}$  appears as the top independent driver of the ceiling $\Hagl$. Its raw Spearman correlation is $\rho_S = +0.44$, $+0.49$, $+0.44$, and $+0.54$ in HMtn, LMtn, Hills, and Plains respectively, with retentions of $0.61$, $0.45$, $0.48$, and $0.72$. In three of the four terrains, $\mathtt{blh}$ therefore keeps roughly half or more of its raw signal after removing the linear $\ttd$ contribution; in Plains, where $\mathtt{blh}$ is already the strongest raw correlate ahead of $\ttd$ itself, retention is the largest. The next block of variables in the $\Hagl$ ranking is dominated by near-surface temperature quantities ($\mathtt{t2m}$, $\mathtt{skt}$, $\mathtt{mn2t}$, $\mathtt{mx2t}$, $\mathtt{stl1}$) and clear-sky longwave fluxes ($\mathtt{avg\_tnlwrfcs}$, $\mathtt{avg\_snlwrfcs}$), all of which start from moderate raw correlations ($|\rho_S| \sim 0.25$--$0.42$) but collapse after orthogonalization, with retentions falling to $0.20$--$0.50$ in the mountains and to $0.05$--$0.20$ in Plains: in the flatter terrain, these variables are essentially proxies for $\ttd$ and carry little information beyond it. For the mean climb rate $\Emvz$, $\mathtt{blh}$ is again the top driver in every terrain, with raw $\rho_S = +0.22$ (HMtn), $+0.34$ (LMtn), $+0.20$ (Hills), and $+0.19$ (Plains) and retentions in the range $0.72$--$0.85$. In HMtn and LMtn the next-strongest independent predictors form a group of wind variables ($\mathtt{v10}$, $\mathtt{v10n}$, $\mathtt{u10}$, $\mathtt{v100}$, $\mathtt{u100}$) with retentions $0.72$--$0.91$; in Hills and Plains this wind block is largely absent and is replaced by soil-moisture ($\mathtt{swvl2}$, $\mathtt{swvl3}$) and shortwave-radiation variables. The vertical-velocity variability $\Esvz$ shows uniformly weak raw correlations ($|\rho_S| \lesssim 0.15$) in all four terrains with no single dominant variable, but retentions of the top-ranked candidates remain high ($0.7$--$0.95$), suggesting that whatever weak signal exists is genuinely independent of $\ttd$.\\

\paragraph*{Cold midday.}
Figure~\ref{fig:orfr_cold} shows the same analysis for the Cold season. The most visible change relative to Warm midday is that many top-ranked variables now exceed $\ttd$ in raw correlation and also survive orthogonalization with high retention. For $\Hagl$, the leading raw correlate is $\mathtt{blh}$ in HMtn ($\rho_S = +0.50$ vs $\mathtt{T\_Td}$ at $+0.27$) and in Plains ($+0.68$ vs $+0.54$), and the skin (surface) temperature $\mathtt{skt}$ in LMtn ($+0.52$ vs $+0.40$) and Hills ($+0.51$ vs $+0.44$). Retentions after orthogonalization are systematically larger than in the Warm season: $\mathtt{blh}$ retains $0.87$, $0.79$, $0.62$, and $0.71$ of its raw signal across the four terrains, and $\mathtt{skt}$ retains $0.79$ (HMtn), $0.75$ (LMtn), and $0.62$ (Hills). This general rise in retention reflects the weaker Cold-season $\ttd$ signal identified in the previous section: because $\ttd$ carries less information about $\Hagl$ in Cold conditions, the orthogonalization step removes less from every candidate variable. Beyond $\mathtt{blh}$ and $\mathtt{skt}$, the ceiling rankings feature the near-surface temperature cluster ($\mathtt{t2m}$, $\mathtt{mn2t}$, $\mathtt{mx2t}$), the soil-temperature profile ($\mathtt{stl1}$--$\mathtt{stl3}$), and clear-sky radiation fluxes ($\mathtt{avg\_tnlwrfcs}$, $\mathtt{avg\_sdlwrfcs}$, $\mathtt{avg\_snswrfcs}$), all with retentions in the $0.60$--$0.90$ range in HMtn, LMtn, and Hills. For $\Emvz$, $\mathtt{blh}$ again heads the ranking in all four terrains, with retentions in the $0.73$--$0.96$ range; shortwave-radiation variables ($\mathtt{avg\_snswrfcs}$, $\mathtt{avg\_tdswrf}$, $\mathtt{avg\_sdswrfcs}$) gain prominence particularly in Hills and Plains, where they cluster near $\mathtt{blh}$ at the top of the list. The vertical-velocity variability $\Esvz$ shows higher $|\rho_S|$ than in the Warm season ($0.15$--$0.25$ instead of $\lesssim 0.15$), with the sensible-heat flux $\mathtt{avg\_ishf}$ and the shortwave-radiation cluster as leading candidates.
The Plains Cold configuration deserves a cautionary note: it displays the highest raw correlations of the entire set, together with an uneven retention pattern ($0.71$ for $\mathtt{blh}$ but only $0.09$ for $\mathtt{skt}$). Understanding this configuration would require additional data and targeted studies, which we leave to future work.

\section{Conclusion}
\label{sec:conc}
Let us summarise what we have achieved. We have introduced a methodology to transform paragliding flight logs into a quantitative dataset for the study of atmospheric thermal convection. From $110{,}730$ IGC flight records over metropolitan France during 2017--2024 we isolated $1.47$ million climbing segments, each of which acts as a probe of one thermal updraft. From every climb we extracted three physically distinct observables that probe the vertical extent, the mean buoyancy generation, or thermal strength, and the small-scale velocity fluctuations of individual updrafts. 

We revealed three robust descriptive features of the observed climatology of thermal convection. First, the terrain hierarchy High mountains $>$ Low mountains $>$ Hills $>$ Plains is preserved across all three observables and all combinations of season, hour, cloud cover, and soil-moisture state, reflecting the effect of orography on convection development. Second, the seasonal and diurnal cycles produce large modulations of the ceiling and mean climb rate but only modest changes in vertical-velocity variability. Third, at fixed terrain and hour, soil moisture is the dominant conditional axis, with dry soil significantly raising the ceiling in Low mountains, whereas cloud cover state contributes a weaker and terrain-dependent modulation.

Combined with a reduced ERA5 catalog of physically interpretable predictors and implemented through an orthogonalized rank-based feature-relevance framework, the same pipeline yields three main results regarding the atmospheric predictors of the observables. First, in Low mountains during the Warm season at midday, the empirical slope of ceiling height with respect to the temperature–dew point deficit matches the theoretical lifting condensation level with striking precision. This match constitutes a first-principles validation of the methodology and paragliding-derived ceiling. 
Second, the orthogonal regression on temperature--dew point-deficit-decorrelated predictors identifies the boundary-layer height as the leading independent driver of both ceiling and thermal strength across all four terrains and both seasons, while the vertical-velocity variability is instead controlled by surface heat-flux and wind variables. Third, the Warm-to-Cold transition displays a systematic increase in the retention of these decorrelated predictors, consistent with the weakening of the lifting condensation level scaling in the cold half of the year, and skin temperature replaces boundary-layer height as the leading raw correlate of the ceiling in Low mountains and Hills.

Let us now discuss some limitations of the present study, and ideas for future work. It should be noted that our dataset  inherits the sampling bias of the free-flying activity: paragliding is only practiced in flyable conditions (moderate winds, sufficient thermal activity, sufficient visibility, etc.). This implies that our observables  are implicitly conditioned on the subset of atmospheric states that admit paragliding flights. The correlations reported here therefore hold within that operational window, not on the full climatological distribution.

Beyond the France 2017--2024 demonstration, the pipeline described here is applicable to any region and time period covered by an open flight-log platform and a reanalysis product. Three natural extensions are worth mentioning. First, the spatio-temporal regularity of thermal columns could be addressed through inter-thermal spacing extracted from the transition phases of the flight tracks. Second, the correlations we obtain for the vertical-velocity variability suggest that small-scale, higher-order moments of the vertical velocity field within individual updrafts encode sub-grid information that is inaccessible to the $\sim$25km reanalysis grid; systematic exploration of these moments will require the segment-level, rather than cell-hour, resolution of the paragliding data and higher-resolution meteorological products. In addition, examining lagged correlations would be a valuable direction for clarifying the temporal relationships between predictors and observables. All three directions should benefit from the larger sample sizes and additional terrain classes that a global extension of the pipeline would enable.

\section{Acknowledgments}
We would like to thank Damien Challet, Alexandre Darmon, Anna Lea Albright, Rohit Kumar Ramesh, and Jérémie Vilpellet for fruitful discussions. 
This research was conducted within the Econophysics \& Complex Systems Research Chair, under the aegis of the Fondation du Risque and Capital Fund Management. To help with editing and overall polishing of the manuscript we used an AI-assisted writing tool, while all substantive ideas, analysis, and conclusions remain the responsibility of the authors.

\bibliography{refs}

\clearpage 

\onecolumngrid

\appendix
\label{AP:AP1}

\section{More summary statistics of conditional splits}
\label{sec:stats_tab}
\begin{table*}[htbp]
\centering
\caption{Summary statistics of thermal observables by terrain, season, and hour block. Values are means over all cell-hour observations.}
\label{tab:stats_seasonal_diurnal}
\renewcommand{\arraystretch}{1.1}
\setlength{\tabcolsep}{5pt}
\small
\begin{tabular}{lllrrrr}
\toprule
Terrain & Season & Hour block & $n$ & $\bar{H}_{\mathrm{AGL}}$ [m] & $\bar{V}_z$ [m s$^{-1}$] & $\delta{V_z}$ [m s$^{-1}$] \\
\midrule
High mountains & Warm & AllDay & 19,746 & 1613 & 1.35 & 1.40 \\
 &  & Morning & 7,119 & 1513 & 1.25 & 1.40 \\
 &  & Midday & 8,648 & 1716 & 1.41 & 1.43 \\
 &  & Afternoon & 3,979 & 1566 & 1.36 & 1.34 \\
 & Cold & AllDay & 4,752 & 1371 & 1.13 & 1.36 \\
 &  & Morning & 1,120 & 1266 & 1.09 & 1.38 \\
 &  & Midday & 3,110 & 1429 & 1.15 & 1.36 \\
 &  & Afternoon & 522 & 1255 & 1.11 & 1.28 \\
\midrule
Low mountains & Warm & AllDay & 58,913 & 1529 & 1.20 & 1.33 \\
 &  & Morning & 17,606 & 1443 & 1.17 & 1.38 \\
 &  & Midday & 29,023 & 1598 & 1.24 & 1.34 \\
 &  & Afternoon & 12,284 & 1488 & 1.13 & 1.25 \\
 & Cold & AllDay & 12,547 & 1199 & 1.01 & 1.27 \\
 &  & Morning & 2,525 & 1160 & 1.05 & 1.34 \\
 &  & Midday & 8,217 & 1221 & 1.03 & 1.28 \\
 &  & Afternoon & 1,805 & 1153 & 0.92 & 1.16 \\
\midrule
Hills & Warm & AllDay & 40,542 & 1444 & 1.12 & 1.23 \\
 &  & Morning & 6,521 & 1406 & 1.18 & 1.31 \\
 &  & Midday & 24,211 & 1492 & 1.14 & 1.24 \\
 &  & Afternoon & 9,810 & 1350 & 1.01 & 1.13 \\
 & Cold & AllDay & 7,268 & 1174 & 0.98 & 1.19 \\
 &  & Morning & 808 & 1226 & 1.08 & 1.28 \\
 &  & Midday & 5,050 & 1195 & 1.00 & 1.20 \\
 &  & Afternoon & 1,410 & 1068 & 0.85 & 1.09 \\
\midrule
Plains & Warm & AllDay & 24,166 & 1313 & 1.01 & 1.16 \\
 &  & Morning & 4,512 & 1151 & 1.05 & 1.22 \\
 &  & Midday & 14,459 & 1353 & 1.03 & 1.17 \\
 &  & Afternoon & 5,195 & 1343 & 0.91 & 1.07 \\
 & Cold & AllDay & 3,805 & 1118 & 0.92 & 1.12 \\
 &  & Morning & 503 & 931 & 0.95 & 1.20 \\
 &  & Midday & 2,646 & 1131 & 0.95 & 1.12 \\
 &  & Afternoon & 656 & 1209 & 0.80 & 1.03 \\
\bottomrule
\end{tabular}
\end{table*}

\begin{table*}[htbp]
\centering
\caption{Cloud and soil-moisture effects on thermal observables at midday (12--14h solar, Warm season).}
\label{tab:stats_cloud_soil}
\renewcommand{\arraystretch}{1.05}
\setlength{\tabcolsep}{5pt}
\small
\begin{tabular}{lllrrrr}
\toprule
Terrain & Cloud & Soil & $n$ & $\bar{H}_{\mathrm{AGL}}$ [m] & $\bar{V}_z$ [m s$^{-1}$] & $\delta{V_z}$ [m s$^{-1}$] \\
\midrule
High mountains & All & All & 8,648 & 1716 & 1.41 & 1.43 \\
 & Clear & All & 4,205 & 1694 & 1.38 & 1.43 \\
 & Cloudy & All & 4,443 & 1737 & 1.44 & 1.43 \\
 & All & Dry & 3,044 & 1935 & 1.53 & 1.47 \\
 & All & Wet & 5,604 & 1597 & 1.35 & 1.41 \\
\cmidrule(l){2-7}
 & Clear & Dry & 1,414 & 1943 & 1.53 & 1.47 \\
 & Clear & Wet & 2,791 & 1568 & 1.31 & 1.40 \\
 & Cloudy & Dry & 1,630 & 1929 & 1.53 & 1.46 \\
 & Cloudy & Wet & 2,813 & 1626 & 1.39 & 1.41 \\
\midrule
Low mountains & All & All & 29,023 & 1598 & 1.24 & 1.34 \\
 & Clear & All & 15,717 & 1626 & 1.21 & 1.34 \\
 & Cloudy & All & 13,306 & 1565 & 1.27 & 1.33 \\
 & All & Dry & 7,851 & 1883 & 1.42 & 1.39 \\
 & All & Wet & 21,172 & 1492 & 1.17 & 1.32 \\
\cmidrule(l){2-7}
 & Clear & Dry & 4,602 & 1920 & 1.39 & 1.40 \\
 & Clear & Wet & 11,115 & 1505 & 1.13 & 1.32 \\
 & Cloudy & Dry & 3,249 & 1831 & 1.44 & 1.38 \\
 & Cloudy & Wet & 10,057 & 1478 & 1.21 & 1.31 \\
\midrule
Hills & All & All & 24,211 & 1492 & 1.14 & 1.24 \\
 & Clear & All & 14,307 & 1502 & 1.13 & 1.25 \\
 & Cloudy & All & 9,904 & 1478 & 1.16 & 1.23 \\
 & All & Dry & 8,635 & 1632 & 1.21 & 1.27 \\
 & All & Wet & 15,576 & 1415 & 1.10 & 1.23 \\
\cmidrule(l){2-7}
 & Clear & Dry & 5,294 & 1629 & 1.20 & 1.28 \\
 & Clear & Wet & 9,013 & 1428 & 1.09 & 1.23 \\
 & Cloudy & Dry & 3,341 & 1638 & 1.23 & 1.26 \\
 & Cloudy & Wet & 6,563 & 1397 & 1.12 & 1.22 \\
\midrule
Plains & All & All & 14,459 & 1353 & 1.03 & 1.17 \\
 & Clear & All & 8,828 & 1321 & 1.00 & 1.17 \\
 & Cloudy & All & 5,631 & 1403 & 1.08 & 1.16 \\
 & All & Dry & 6,521 & 1440 & 1.08 & 1.19 \\
 & All & Wet & 7,938 & 1281 & 0.99 & 1.15 \\
\cmidrule(l){2-7}
 & Clear & Dry & 4,125 & 1407 & 1.05 & 1.19 \\
 & Clear & Wet & 4,703 & 1246 & 0.96 & 1.15 \\
 & Cloudy & Dry & 2,396 & 1498 & 1.14 & 1.19 \\
 & Cloudy & Wet & 3,235 & 1333 & 1.04 & 1.14 \\
\bottomrule
\end{tabular}
\end{table*}

\FloatBarrier 

\clearpage

\section{ERA5 predictors used in the ORFR analysis}
\label{ap:vars}
\onecolumngrid
\tablefirsthead{%
\toprule
\textbf{Short name} & \textbf{Description} & \textbf{Units}\\
\midrule}
\tablehead{%
\multicolumn{3}{c}{\footnotesize\textit{(continued from previous page)}}\\
\toprule
\textbf{Short name} & \textbf{Description} & \textbf{Units}\\
\midrule}
\tabletail{%
\midrule
\multicolumn{3}{r}{\footnotesize\textit{(continued on next page)}}\\}
\tablelasttail{\bottomrule}

\bottomcaption{The 120 ERA5 predictors used in the orthogonal regressionanalysis of Section~\ref{sec:orfr}, grouped by physical family. Names anddescriptions follow the ECMWF ERA5 catalogue as retrieved through the CDS API. Time-averaged quantities are prefixed $\mathtt{avg\_}$. Temperatures originally provided in K by ERA5 are converted to $^\circ$C for the analysis. The dew point depression $T{-}T_d$ and the wind speeds $\mathtt{uv10}$, $\mathtt{uv100}$ are derived quantities computed from the raw variables. Abbreviations: CS = clear-sky; VI = vertically integrated; PP = previous post-processing.\label{tab:era5_predictors}}

\begin{center}\small
\begin{supertabular}{@{}lp{6.8cm}l@{}}

\multicolumn{3}{@{}l}{\textit{Near-surface temperature}}\\
$\mathtt{skt}$   & Skin temperature                                          & $^\circ$C\\
$\mathtt{t2m}$   & 2\,m air temperature                                      & $^\circ$C\\
$\mathtt{mn2t}$  & Minimum 2\,m temperature since previous PP                & $^\circ$C\\
$\mathtt{mx2t}$  & Maximum 2\,m temperature since previous PP                & $^\circ$C\\
\midrule

\multicolumn{3}{@{}l}{\textit{Soil temperature}}\\
$\mathtt{stl1}$  & Soil temperature, layer 1 (0--7\,cm)                      & $^\circ$C\\
$\mathtt{stl2}$  & Soil temperature, layer 2 (7--28\,cm)                     & $^\circ$C\\
$\mathtt{stl3}$  & Soil temperature, layer 3 (28--100\,cm)                   & $^\circ$C\\
$\mathtt{stl4}$  & Soil temperature, layer 4 (100--289\,cm)                  & $^\circ$C\\
\midrule

\multicolumn{3}{@{}l}{\textit{Soil moisture}}\\
$\mathtt{swvl2}$ & Volumetric soil water, layer 2 (7--28\,cm)                & m$^3$\,m$^{-3}$\\
$\mathtt{swvl3}$ & Volumetric soil water, layer 3 (28--100\,cm)              & m$^3$\,m$^{-3}$\\
$\mathtt{swvl4}$ & Volumetric soil water, layer 4 (100--289\,cm)             & m$^3$\,m$^{-3}$\\
\midrule

\multicolumn{3}{@{}l}{\textit{Moisture and cloud (column-integrated)}}\\
$\mathtt{tcw}$   & Total column water                                        & kg\,m$^{-2}$\\
$\mathtt{tcwv}$  & Total column water vapour                                 & kg\,m$^{-2}$\\
$\mathtt{tciw}$  & Total column cloud ice water                              & kg\,m$^{-2}$\\
$\mathtt{tclw}$  & Total column cloud liquid water                           & kg\,m$^{-2}$\\
$\mathtt{tcslw}$ & Total column supercooled liquid water                     & kg\,m$^{-2}$\\
$\mathtt{lcc}$   & Low cloud cover                                           & (0--1)\\
$\mathtt{hcc}$   & High cloud cover                                          & (0--1)\\
\midrule

\multicolumn{3}{@{}l}{\textit{Boundary-layer and vertical structure}}\\
$\mathtt{blh}$       & Boundary-layer height                                 & m\\
$\mathtt{cbh}$       & Cloud base height                                     & m\\
$\mathtt{deg0l}$     & Zero-degree isotherm height                           & m\\
$\mathtt{avg\_ibld}$ & Mean boundary-layer dissipation                       & W\,m$^{-2}$\\
$\mathtt{dctb}$      & Duct base height                                      & m\\
$\mathtt{tplt}$      & Trapping-layer top height                             & m\\
$\mathtt{tplb}$      & Trapping-layer base height                            & m\\
$\mathtt{dndza}$     & Mean refractivity gradient in trapping layer          & m$^{-1}$\\
$\mathtt{dndzn}$     & Minimum refractivity gradient in trapping layer       & m$^{-1}$\\
\midrule

\multicolumn{3}{@{}l}{\textit{Radiation --- shortwave (time-mean fluxes)}}\\
$\mathtt{avg\_sdswrf}$     & Mean surface downward SW radiation flux         & W\,m$^{-2}$\\
$\mathtt{avg\_snswrf}$     & Mean surface net SW radiation flux              & W\,m$^{-2}$\\
$\mathtt{avg\_sdirswrf}$   & Mean surface direct SW radiation flux           & W\,m$^{-2}$\\
$\mathtt{avg\_sduvrf}$     & Mean surface downward UV radiation flux         & W\,m$^{-2}$\\
$\mathtt{avg\_tnswrf}$     & Mean top-of-atmosphere net SW radiation flux    & W\,m$^{-2}$\\
$\mathtt{avg\_tdswrf}$     & Mean top-of-atmosphere downward SW radiation    & W\,m$^{-2}$\\
$\mathtt{avg\_sdswrfcs}$   & Mean surface downward SW flux (CS)              & W\,m$^{-2}$\\
$\mathtt{avg\_snswrfcs}$   & Mean surface net SW flux (CS)                   & W\,m$^{-2}$\\
$\mathtt{avg\_sdirswrfcs}$ & Mean surface direct SW flux (CS)                & W\,m$^{-2}$\\
$\mathtt{avg\_tnswrfcs}$   & Mean top-of-atmosphere net SW flux (CS)         & W\,m$^{-2}$\\
\midrule

\multicolumn{3}{@{}l}{\textit{Radiation --- longwave (time-mean fluxes)}}\\
$\mathtt{avg\_sdlwrf}$   & Mean surface downward LW radiation flux           & W\,m$^{-2}$\\
$\mathtt{avg\_snlwrf}$   & Mean surface net LW radiation flux                & W\,m$^{-2}$\\
$\mathtt{avg\_tnlwrf}$   & Mean top-of-atmosphere net LW radiation flux      & W\,m$^{-2}$\\
$\mathtt{avg\_sdlwrfcs}$ & Mean surface downward LW flux (CS)                & W\,m$^{-2}$\\
$\mathtt{avg\_snlwrfcs}$ & Mean surface net LW flux (CS)                     & W\,m$^{-2}$\\
$\mathtt{avg\_tnlwrfcs}$ & Mean top-of-atmosphere net LW flux (CS)           & W\,m$^{-2}$\\
\midrule

\multicolumn{3}{@{}l}{\textit{Surface energy and moisture fluxes (time-mean)}}\\
$\mathtt{avg\_ishf}$    & Mean surface sensible heat flux                    & W\,m$^{-2}$\\
$\mathtt{avg\_slhtf}$   & Mean surface latent heat flux                      & W\,m$^{-2}$\\
$\mathtt{avg\_ie}$      & Mean evaporation rate                              & kg\,m$^{-2}$\,s$^{-1}$\\
$\mathtt{avg\_pevr}$    & Mean potential evaporation rate                    & kg\,m$^{-2}$\,s$^{-1}$\\
$\mathtt{avg\_iews}$    & Mean eastward turbulent surface stress             & N\,m$^{-2}$\\
$\mathtt{avg\_inss}$    & Mean northward turbulent surface stress            & N\,m$^{-2}$\\
$\mathtt{avg\_iegwss}$  & Mean eastward gravity-wave surface stress          & N\,m$^{-2}$\\
$\mathtt{avg\_ingwss}$  & Mean northward gravity-wave surface stress         & N\,m$^{-2}$\\
$\mathtt{avg\_igwd}$    & Mean gravity-wave dissipation                      & W\,m$^{-2}$\\
$\mathtt{lgws}$         & Eastward gravity-wave surface stress               & N\,m$^{-2}$\\
$\mathtt{mgws}$         & Northward gravity-wave surface stress              & N\,m$^{-2}$\\
$\mathtt{zust}$         & Friction velocity                                  & m\,s$^{-1}$\\
\midrule

\multicolumn{3}{@{}l}{\textit{Wind (10\,m, 100\,m, gust)}}\\
$\mathtt{u10}$    & 10\,m eastward wind component                             & m\,s$^{-1}$\\
$\mathtt{v10}$    & 10\,m northward wind component                            & m\,s$^{-1}$\\
$\mathtt{u10n}$   & 10\,m neutral eastward wind                               & m\,s$^{-1}$\\
$\mathtt{v10n}$   & 10\,m neutral northward wind                              & m\,s$^{-1}$\\
$\mathtt{u100}$   & 100\,m eastward wind component                            & m\,s$^{-1}$\\
$\mathtt{v100}$   & 100\,m northward wind component                           & m\,s$^{-1}$\\
$\mathtt{fg10}$   & 10\,m wind gust since previous PP                         & m\,s$^{-1}$\\
$\mathtt{i10fg}$  & Instantaneous 10\,m wind gust                             & m\,s$^{-1}$\\
\midrule

\multicolumn{3}{@{}l}{\textit{Vertically integrated scalar quantities}}\\
$\mathtt{vima}$   & VI mass of the atmosphere                                 & kg\,m$^{-2}$\\
$\mathtt{vimat}$  & VI mass tendency                                          & kg\,m$^{-2}$\,s$^{-1}$\\
$\mathtt{vike}$   & VI kinetic energy                                         & J\,m$^{-2}$\\
$\mathtt{vithe}$  & VI thermal energy                                         & J\,m$^{-2}$\\
$\mathtt{vitoe}$  & VI total energy                                           & J\,m$^{-2}$\\
$\mathtt{vit}$    & VI temperature                                            & K\,kg\,m$^{-2}$\\
$\mathtt{vipie}$  & VI potential and internal energy                          & J\,m$^{-2}$\\
$\mathtt{vipile}$ & VI potential, internal and latent energy                  & J\,m$^{-2}$\\
$\mathtt{viec}$   & VI energy conversion                                      & W\,m$^{-2}$\\
\midrule

\multicolumn{3}{@{}l}{\textit{Vertically integrated divergences}}\\
$\mathtt{vimd}$       & VI moisture divergence                                & kg\,m$^{-2}$\\
$\mathtt{avg\_vimdf}$ & Mean VI moisture-flux divergence                      & kg\,m$^{-2}$\,s$^{-1}$\\
$\mathtt{vimdf}$      & VI moisture-flux divergence                           & kg\,m$^{-2}$\,s$^{-1}$\\
$\mathtt{vimad}$      & VI mass-flux divergence                               & kg\,m$^{-2}$\,s$^{-1}$\\
$\mathtt{vited}$      & VI total-energy-flux divergence                       & W\,m$^{-2}$\\
$\mathtt{viked}$      & VI kinetic-energy-flux divergence                     & W\,m$^{-2}$\\
$\mathtt{vilwd}$      & VI cloud-liquid-water-flux divergence                 & kg\,m$^{-2}$\,s$^{-1}$\\
$\mathtt{viiwd}$      & VI cloud-frozen-water-flux divergence                 & kg\,m$^{-2}$\,s$^{-1}$\\
$\mathtt{viozd}$      & VI ozone-flux divergence                              & kg\,m$^{-2}$\,s$^{-1}$\\
$\mathtt{vigd}$       & VI geopotential-flux divergence                       & W\,m$^{-2}$\\
$\mathtt{vithed}$     & VI thermal-energy-flux divergence                     & W\,m$^{-2}$\\
\midrule

\multicolumn{3}{@{}l}{\textit{Vertically integrated eastward fluxes}}\\
$\mathtt{vimae}$  & VI eastward mass flux                                     & kg\,m$^{-1}$\,s$^{-1}$\\
$\mathtt{vige}$   & VI eastward geopotential flux                             & W\,m$^{-1}$\\
$\mathtt{vikee}$  & VI eastward kinetic-energy flux                           & W\,m$^{-1}$\\
$\mathtt{vithee}$ & VI eastward heat flux                                     & W\,m$^{-1}$\\
$\mathtt{vitee}$  & VI eastward total-energy flux                             & W\,m$^{-1}$\\
$\mathtt{viwve}$  & VI eastward water-vapour flux                             & kg\,m$^{-1}$\,s$^{-1}$\\
$\mathtt{vilwe}$  & VI eastward cloud-liquid-water flux                       & kg\,m$^{-1}$\,s$^{-1}$\\
$\mathtt{viiwe}$  & VI eastward cloud-frozen-water flux                       & kg\,m$^{-1}$\,s$^{-1}$\\
$\mathtt{vioze}$  & VI eastward ozone flux                                    & kg\,m$^{-1}$\,s$^{-1}$\\
\midrule

\multicolumn{3}{@{}l}{\textit{Vertically integrated northward fluxes}}\\
$\mathtt{viman}$  & VI northward mass flux                                    & kg\,m$^{-1}$\,s$^{-1}$\\
$\mathtt{vign}$   & VI northward geopotential flux                            & W\,m$^{-1}$\\
$\mathtt{viken}$  & VI northward kinetic-energy flux                          & W\,m$^{-1}$\,s$^{-1}$\\
$\mathtt{vithen}$ & VI northward heat flux                                    & W\,m$^{-1}$\\
$\mathtt{viten}$  & VI northward total-energy flux                            & W\,m$^{-1}$\\
$\mathtt{viwvn}$  & VI northward water-vapour flux                            & kg\,m$^{-1}$\,s$^{-1}$\\
$\mathtt{vilwn}$  & VI northward cloud-liquid-water flux                      & kg\,m$^{-1}$\,s$^{-1}$\\
$\mathtt{viiwn}$  & VI northward cloud-frozen-water flux                      & kg\,m$^{-1}$\,s$^{-1}$\\
$\mathtt{viozn}$  & VI northward ozone flux                                   & kg\,m$^{-1}$\,s$^{-1}$\\
\midrule

\multicolumn{3}{@{}l}{\textit{Instability indices}}\\
$\mathtt{kx}$     & K-index                                                   & K\\
$\mathtt{totalx}$ & Total totals index                                        & K\\
\midrule

\multicolumn{3}{@{}l}{\textit{Surface characteristics}}\\
$\mathtt{fal}$     & Forecast albedo                                          & (0--1)\\
$\mathtt{fsr}$     & Forecast surface roughness                               & m\\
$\mathtt{alnid}$   & Near-IR albedo, diffuse radiation                        & (0--1)\\
$\mathtt{alnip}$   & Near-IR albedo, direct radiation                         & (0--1)\\
$\mathtt{aluvd}$   & UV-visible albedo, diffuse radiation                     & (0--1)\\
$\mathtt{aluvp}$   & UV-visible albedo, direct radiation                      & (0--1)\\
$\mathtt{lai\_lv}$ & Leaf area index, low vegetation                          & m$^2$\,m$^{-2}$\\
$\mathtt{lai\_hv}$ & Leaf area index, high vegetation                         & m$^2$\,m$^{-2}$\\
$\mathtt{src}$     & Skin reservoir water content                             & kg\,m$^{-2}$\\
\midrule

\multicolumn{3}{@{}l}{\textit{Pressure, runoff and snow evaporation}}\\
$\mathtt{sp}$            & Surface pressure                                   & Pa\\
$\mathtt{msl}$           & Mean sea-level pressure                            & Pa\\
$\mathtt{avg\_rorwe}$    & Mean runoff rate                                   & kg\,m$^{-2}$\,s$^{-1}$\\
$\mathtt{avg\_ssurfror}$ & Mean sub-surface runoff rate                       & kg\,m$^{-2}$\,s$^{-1}$\\
$\mathtt{avg\_esrwe}$    & Mean snow evaporation rate                         & kg\,m$^{-2}$\,s$^{-1}$\\
\midrule

\multicolumn{3}{@{}l}{\textit{Derived quantities}}\\
$\mathtt{T\_Td}$ & Dew point depression, $\mathtt{t2m} - \mathtt{d2m}$                            & $^\circ$C\\
$\mathtt{uv10}$   & 10\,m wind speed, $\sqrt{u_{10}^2+v_{10}^2}$     & m\,s$^{-1}$\\
$\mathtt{uv100}$  & 100\,m wind speed, $\sqrt{u_{100}^2+v_{100}^2}$                               & m\,s$^{-1}$\\

\end{supertabular}
\end{center}
\twocolumngrid

\end{document}